\documentclass[a4paper,11pt]{article}
\pdfoutput=1 

\usepackage{jcappub} 

\usepackage[T1]{fontenc} 

\usepackage{graphicx}

\title{Gravitational waveforms from periodic orbits around Gauss-Bonnet black holes}

\author[a]{Yi-Han Huang,}
\author[a,1]{Sen Guo,\note{Corresponding author.}}
\author[b]{Yu Liang,}
\author[a]{Lin Wen,}
\author[c]{Kai Lin,}

\affiliation[a]{College of Physics and Electronic Engineering, Chongqing Normal University, Chongqing 401331, People's Republic of China}
\affiliation[b]{School of Big Data and Artificial Intelligence, Fuyang University of Technology, Fuyang 236000, People's Republic of China}
\affiliation[c]{Universidade Federal de Campina Grande, Campina Grande, PB, Brasil, \\ Instituto de F\'isica, Universidade de S\~ao Paulo, S\~ao Paulo, Brasil}

\emailAdd{sguophys@126.com}

\abstract{Extreme mass-ratio inspirals (EMRIs) constitute one of the most promising probes of strong field gravity for future space borne gravitational-wave observatories. As a representative higher-curvature extension of General Relativity (GR), four-dimensional Einstein-Gauss-Bonnet (4D EGB) gravity is distinguished by its strictly linear geometric coupling. By this mathematical property, the pathological Fisher-matrix singularities that typically plague conventional modified black hole models are effectively evaded, thereby providing an ideal framework to test topological deviations from classical spacetimes. Through the classification of equatorial periodic orbits via an integer taxonomy $(z,w,v)$, it is demonstrated that even modest Gauss-Bonnet couplings ($\alpha \sim 0.1M^2$) imprint measurable geometric signatures onto the zoom-whirl architecture. Although the global conservative energy budget is shifted by a mere $\sim 0.2\%$, the short-range repulsive EGB core severely alters the strong field whirl dynamics, whereby a resolvable macroscopic dephasing of several radians per orbit is accumulated. Through semi-relativistic waveform modeling, it is revealed that this temporal compression manifests as a rigid, high-frequency stretching of the gravitational-wave harmonic comb---a clean, amplitude-independent spectral signature ideally suited for detection by LISA, Taiji, and TianQin. A rigorous Fisher information analysis confirms that for a typical four-year observation at a signal-to-noise ratio of $\rho=20$, the marginalized error on the EGB coupling can be tightly bounded to $\sigma_\alpha \sim \mathcal{O}(10^{-6}) M^2$, with virtually negligible parameter degeneracy with the orbital eccentricity. Furthermore, by incorporating contemporary black hole shadow measurements, a highly complementary multi-messenger framework is established, bridging the dynamical timelike EMRI observables with the static null geodesic sector. Ultimately, it is established that the strong-field phase evolution inherent to zoom-whirl EMRIs provides a remarkably robust avenue for constraining higher-curvature quantum corrections beyond GR.}

\begin{document}
\maketitle
\flushbottom

\section{Introduction}
\label{sec:intro}
\par
Although General Relativity (GR) has been corroborated by extensive precision observational tests, the inclusion of higher-curvature corrections is strongly motivated by the inevitable emergence of spacetime singularities and the absence of a renormalizable quantum gravity theory~\cite{Will:2014kxa}. Among these theoretical extensions, particular physical significance is attributed to the Einstein-Gauss-Bonnet (EGB) model, wherein the Gauss-Bonnet invariant naturally arises as the leading-order $\alpha'$-correction within the low-energy limit of heterotic string theory \cite{Zwiebach:1985uq}. Although the original mathematical regularization of the four-dimensional EGB model was subjected to theoretical scrutiny, it has been rigorously verified that consistent alternative derivations yield an identical spacetime geometry~\cite{Gurses:2020ofy}. By introducing a well-defined, linear $\mathcal{O}(\alpha)$ geometrical modification to the Schwarzschild spacetime, a theoretical framework is provided through which the macroscopic effects of short-range modified gravity can be quantitatively parameterized.

\par
The observational testing of such strong-field geometrical modifications is fundamentally predicated upon Extreme Mass Ratio Inspirals (EMRIs), which have been designated as primary astrophysical targets for the forthcoming Laser Interferometer Space Antenna (LISA)~\cite{Amaro-Seoane:2012aqc}. However, the generation of full EMRI waveforms in non-Kerr spacetimes remains computationally prohibitive. This bottleneck is primarily attributed to the mathematical complexity inherent in evaluating the gravitational self-force and the corresponding radiation reaction over approximately $\mathcal{O}(10^5)$ orbital cycles~\cite{Barack:2018yvs}. To circumvent this computational limitation, a robust semi-analytical framework is extensively adopted in the literature, which relies upon the analysis of the conservative dynamics associated with equatorial rational periodic orbits~\cite{Levin:2008mq}. By these trajectories, which are characterized by their striking ``zoom-whirl'' behavior, the fine structure of the effective potential near the event horizon is intricately mapped \cite{Kagohashi:2024qkp,Glampedakis:2002ya}. Through the classification of these periodic orbits via an integer taxonomy $(z,w,v)$, fundamental orbital frequencies and proxy gravitational-wave (GW) spectra can be accurately extracted~\cite{Lin:2020wnp}. Serving as a powerful diagnostic tool, this methodology enables the dominant phenomenological imprints of alternative gravities to be evaluated without the necessity of simulating full dissipative inspiral evolutions~\cite{Qiao:2024gfb}.

\par
Consequently, the zoom-whirl periodic-orbit taxonomy has been extensively deployed as a robust theoretical testbed by which the dynamical imprints of non-standard geometries can be probed~\cite{Bambhaniya:2020zno}. This elegant framework has been successfully applied to comprehensively map the strong-field parameter spaces of a diverse array of modified spacetimes, in which configurations ranging from classical charged black holes and naked singularities to more exotic geometries are included~\cite{Rana:2019bsn,Babar:2017gsg,Liu:2018vea,Chan:2025ocy,Yao:2023ziq,Lin:2023eyd,Wei:2019zdf,Azreg-Ainou:2020bfl,Zhao:2024exh,Alloqulov:2025ucf}. Moving beyond purely conservative dynamics, the proxy gravitational-wave fluxes and waveform modulations emitted during these periodic trajectories across various alternative gravity models have been further evaluated in a substantial body of recent literature~\cite{Haroon:2025rzx,Tu:2023xab,Yang:2024lmj,Junior:2024tmi,Li:2024tld,QiQi:2024dwc}. By these comprehensive investigations, the periodic-orbit methodology is firmly established not merely as a theoretical curiosity, but as an indispensable diagnostic baseline for distinguishing beyond-GR features in future EMRI observations.

\par
It should be noted that many phenomenological metrics suffer from a formal breakdown at the GR limit; specifically, if the geometrical correction depends quadratically on the deviation parameter, the derivative of the waveform with respect to this parameter is forced to vanish identically as $g \to 0$. By this mathematical degeneracy, the standard Fisher Information Matrix is fundamentally caused to become singular, whereby conventional linear error analysis is rendered invalid~\cite{Vallisneri:2007ev}. In this work, the established periodic-orbit machinery is integrated with the 4D EGB spacetime, by which a critical mathematical issue frequently encountered in modified gravity parameter estimation is successfully resolved. Furthermore, a strictly linear $\mathcal{O}(\alpha)$ coupling is introduced by the 4D-EGB theory. To computationally exploit this advantage, a ``cancellation-free'' algebraic representation of the metric is implemented, whereby the catastrophic numerical round-off errors that typically occur at infinitesimally small $\alpha$ are completely eliminated. By this implementation, it is theoretically guaranteed that the Fisher matrix remains perfectly regular and well-conditioned exactly at the GR limit. Furthermore, by evaluating both the timelike periodic orbits (which govern EMRI GW emissions) and the null geodesic sector (which dictates black hole shadows), a comprehensive multi-messenger framework is constructed. Through this framework, the EGB coupling $\alpha$ can be tightly constrained by utilizing both projected LISA sensitivities and current Event Horizon Telescope (EHT) observations, thereby allowing higher-curvature gravity to be tested with a precision that far exceeds that of traditional electromagnetic methods \cite{EventHorizonTelescope:2019dse,Zeng:2020dco}.

\par
The structure of this paper is organized as follows. In Section 2, the geodesic dynamics of the 4D EGB spacetime are derived, the allowed bound-orbit parameter space is mapped, and the zoom-whirl topologies are classified via the $(z,w,v)$ integer taxonomy. In Section 3, the associated gravitational-wave signatures are investigated, after which Fisher-matrix forecasts for space-based interferometers are performed. A complementary multi-messenger probe incorporating EHT shadow constraints is subsequently constructed, and the analytical framework is extended to rotating Kerr spacetimes. Finally, our conclusions are summarized and future prospects are discussed in Section 4.

\section{Geodesic dynamics and periodic orbits in Gauss-Bonnet gravity}
\label{sec:2}
\par
In this section, the static, spherically symmetric black hole solution in 4D Einstein-Gauss-Bonnet (EGB) gravity is considered. Although the Gauss-Bonnet invariant $\mathcal{G} \equiv R^2 - 4 R_{\mu\nu}R^{\mu\nu} + R_{\mu\nu\rho\sigma}R^{\mu\nu\rho\sigma}$ constitutes a strictly topological surface term in exactly four dimensions, a non-trivial, finite contribution to the gravitational dynamics is yielded when the coupling constant is rescaled as $\alpha \to \alpha/(D-4)$ and the singular limit $D \to 4$ is taken~\cite{Torii:2005xu}. The modified action in $D$ dimensions is given by
\begin{equation}
\label{eq-1}
S = \frac{1}{16\pi} \int d^D x \sqrt{-g} \left( R + \frac{\alpha}{D-4} \mathcal{G} \right).
\end{equation}
Geometric units ($G = c = M = 1$) are adopted, by which all physical quantities, including the coupling parameter $\alpha$, are rendered dimensionless. In the $D \to 4$ limit, the static and spherically symmetric vacuum metric is expressed as
\begin{equation}
\label{eq-2}
ds^2 = -f(r)dt^2 + \frac{1}{f(r)}dr^2 + r^2\left(d\theta^2 + \sin^2\theta d\phi^2\right),
\end{equation}
where the lapse function $f(r)$ is given by
\begin{equation}
\label{eq-3}
f(r) = 1 + \frac{r^2}{2\alpha} \left[ 1 - \sqrt{1 + \frac{8\alpha}{r^3}} \right].
\end{equation}
In the weak-field limit ($r \to \infty$), the metric function can be expanded as~\cite{Torii:2005nh}
\begin{equation}
\label{eq-4}
f(r) \simeq 1 - \frac{2}{r} + \frac{4\alpha}{r^4} + \mathcal{O}(r^{-7}).
\end{equation}
By this asymptotic behavior, asymptotic flatness ($\lim_{r \to \infty} f(r) = 1$) is ensured, and the classical Schwarzschild solution is recovered at zeroth order. Because the leading-order higher-curvature correction scales as $\mathcal{O}(\alpha/r^4)$, it is guaranteed that deviations from GR are strongly suppressed at large distances and are tightly localized to the strong-field vicinity of the black hole. The horizon radii are determined by the roots of $f(r) = 0$. For $\alpha > 0$, this condition is simplified to the quadratic equation $r^2 - 2r + \alpha = 0$, by which the outer (event) horizon $r_+$ and the inner (Cauchy) horizon $r_-$ are yielded, i.e.,
\begin{equation}
\label{eq-5}
r_\pm = 1 \pm \sqrt{1 - \alpha}.
\end{equation}
To prevent naked singularities and respect the weak cosmic censorship conjecture, the coupling parameter must be restricted to the physical domain $0 \leq \alpha \leq 1$. The Schwarzschild black hole is recovered in the limit $\alpha = 0$, whereas the extremal configuration, wherein the horizons become degenerate ($r_+ = r_-$), is defined by $\alpha = 1$. Within this allowed parameter space, it is observed that a monotonic decrease in the event horizon radius $r_+$ is induced as the EGB coupling $\alpha$ is increased.

\par
The dynamics of a massive test particle are governed by the Lagrangian
\begin{equation}
\label{eq-6}
2\mathcal{L} = g_{\mu\nu}\dot{x}^\mu \dot{x}^\nu = -f(r)\dot{t}^2 + \frac{\dot{r}^2}{f(r)} + r^2\left(\dot{\theta}^2 + \sin^2\theta\dot{\phi}^2\right),
\end{equation}
wherein the overdot denotes differentiation with respect to proper time $\tau$. Without loss of generality, the motion is restricted to the equatorial plane ($\theta = \pi/2$). Due to the spacetime isometries associated with the Killing vector fields $\partial_t$ and $\partial_\phi$, two conserved quantities along the geodesics are yielded: the specific energy $E$ and specific angular momentum $L$~\cite{Gralla:2019drh},
\begin{equation}
\label{eq-7}
E = f(r)\dot{t}, \qquad L = r^2\dot{\phi}.
\end{equation}
When the timelike normalization condition $2\mathcal{L} = -1$ is imposed and Eq.~\eqref{eq-7} is substituted, the radial equation of motion is derived as
\begin{equation}
\label{eq-8}
\dot{r}^2 = E^2 - V_{\rm eff}(r),
\end{equation}
where the effective potential is defined as
\begin{equation}
\label{eq-9}
V_{\rm eff}(r) = f(r)\left(1 + \frac{L^2}{r^2}\right).
\end{equation}
Bound orbits are permitted exclusively when the particle is trapped within the potential well, a state which necessitates the energy condition $V_{\rm eff}(r) \leq E^2 < 1$.

\par
The existence of circular orbits requires the simultaneous vanishing of both radial velocity ($\dot{r} = 0$) and acceleration ($\ddot{r} = 0$), which is equivalent to the conditions $V_{\rm eff}(r) = E^2$ and $V_{\rm eff}'(r) = 0$, wherein the prime denotes differentiation with respect to $r$. By solving this system, the specific angular momentum and energy for circular orbits at a given radius $r$ are obtained~\cite{Bardeen:1972fi}
\begin{equation}
\label{eq-10}
L_{\rm circ}^2(r) = \frac{r^3 f'(r)}{2f(r) - r f'(r)},
\end{equation}
\begin{equation}
\label{eq-11}
E_{\rm circ}^2(r) = \frac{2f^2(r)}{2f(r) - r f'(r)}.
\end{equation}
The stability of these circular orbits is dictated by the sign of $V_{\rm eff}''(r)$. The innermost stable circular orbit (ISCO), by which the transition between stable circular motion and plunging trajectories is marked, is located at the inflection point $V_{\rm eff}''(r) = 0$. Using Eq.~\eqref{eq-10}, the ISCO radius $r_{\rm ISCO}$ is determined by the roots of
\begin{equation}
\label{eq-12}
r \left[ 2f'(r)^2 - f(r)f''(r) \right] - 3f(r)f'(r) = 0.
\end{equation}
The marginally bound orbit (MBO) is defined as the innermost unstable circular orbit from which precisely the threshold energy required to escape to infinity is possessed by a perturbed particle, corresponding to $E_{\rm circ}^2(r) = 1$. As deduced from Eq.~\eqref{eq-11}, the MBO radius satisfies $r f'(r) = 2f(r)[1 - f(r)]$.

\begin{figure*}[htbp]
\centering
\includegraphics[width=\textwidth]{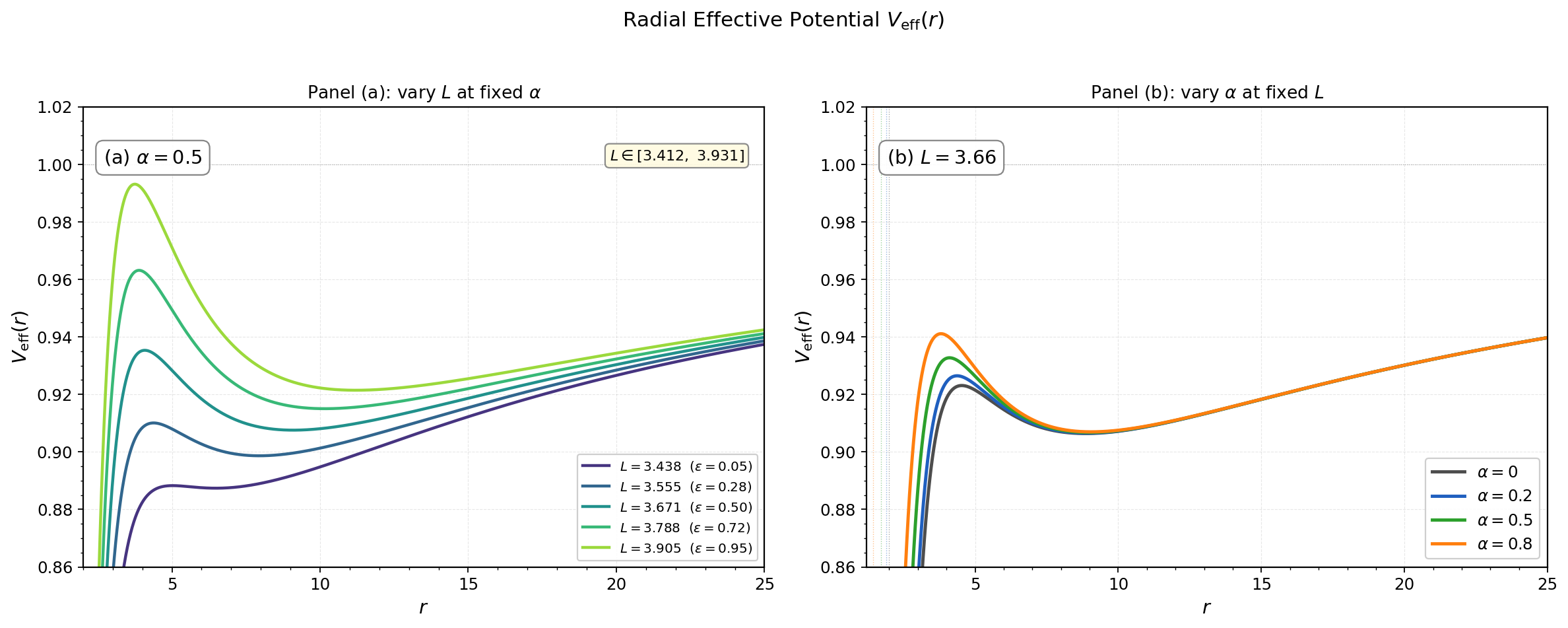}
\caption{Radial effective potential $V_{\rm eff}(r)$ for timelike geodesics. (a) Variation with specific angular momentum $L$ at a fixed EGB coupling $\alpha=0.5$. The dimensionless scaling is defined as $\varepsilon \equiv (L - L_{\rm ISCO})/(L_{\rm MBO} - L_{\rm ISCO})$. (b) Variation with $\alpha$ at a fixed $L=3.66$. By the higher-curvature EGB corrections, the centrifugal barrier is elevated and the event horizon is shrunken (vertical dotted lines), while the potential rapidly degenerates to the Schwarzschild limit ($\alpha=0$) for $r \gtrsim 10$.}
\label{fig:1}
\end{figure*}

\par
For an EMRI system, it is fundamentally understood that the prolonged orbital evolution of the smaller compact object is comprised of bound, precessing trajectories~\cite{Levin:2008mq}, and that the upper limit for the inspiral evolution prior to the plunge is provided by the MBO radius. To systematically investigate the relativistic zoom-whirl orbits that are bounded entirely by the centrifugal barrier, our focus is restricted to the angular momentum domain $L_{\rm ISCO} \leq L \leq L_{\rm MBO}$. The fractional angular momentum $\varepsilon$ is introduced as a dimensionless scaling parameter by which the full dynamically allowed domain of bound orbits is covered. Values over the interval $[0,1]$ are taken by this quantity, where $\varepsilon=0$ corresponds to $L=L_{\rm ISCO}$ and $\varepsilon=1$ corresponds to $L=L_{\rm MBO}$. The dependence of $V_{\rm eff}(r)$ on $L$ and $\alpha$ is illustrated in Fig.~\ref{fig:1}. As $L$ is increased toward $L_{\rm MBO}$, it is observed that the potential well is broadened while the critical escape threshold is approached by the centrifugal barrier. It is confirmed by Panel (b) that the 4D EGB geometrical imprints are strictly localized within the extreme strong-gravity vicinity of the black hole. These periodic orbits are strictly confined to a specific parameter space in the $(L, E)$ plane. When the radial coordinate $r$ is treated as a parameter, the dynamically allowed region is found to be enveloped by four characteristic boundaries:

\textit{Unstable branch:} The upper energy limit is dictated by the unstable circular orbits located at the local maximum of the effective potential. This branch is parameterized by~\cite{Shapiro:1983du}
\begin{equation}
\label{eq-13}
E_{\rm unst}(L) = E_{\rm circ}(r), \quad \text{for} \quad r \in [r_{\rm MBO}, r_{\rm ISCO}].
\end{equation}
Particles possessing energy that exceeds $E_{\rm unst}(L)$ will overcome the barrier and plunge into the black hole.

\textit{Stable branch:} The lower energy limit corresponds to the stable circular orbits located at the local minimum of the potential. This branch is governed by~\cite{Shapiro:1983du}
\begin{equation}
\label{eq-14}
E_{\rm stab}(L) = E_{\rm circ}(r), \quad \text{for} \quad r \in [r_{\rm ISCO}, r_\star],
\end{equation}
wherein $r_\star$ represents the apoapsis radius sharing the identical angular momentum as the MBO, which is numerically found by solving
\begin{equation}
\label{eq-15}
L_{\rm circ}(r_\star) = L_{\rm MBO} \quad (r_\star > r_{\rm ISCO}).
\end{equation}

\textit{ISCO degeneracy and Escape threshold:} The upper and lower branches are smoothly merged at the ISCO $(L_{\rm ISCO}, E_{\rm ISCO})$, by which the absolute minimum angular momentum required for bound motion is marked. Conversely, the allowed domain is abruptly truncated on the right by a vertical segment at $L = L_{\rm MBO}$, whereby the maximum bound energy $E=1$ is connected down to the corresponding stable circular orbit energy $E_\star \equiv E_{\rm circ}(r_\star)$.

\begin{figure}[htbp]
\centering
\includegraphics[width=10cm,height=8cm]{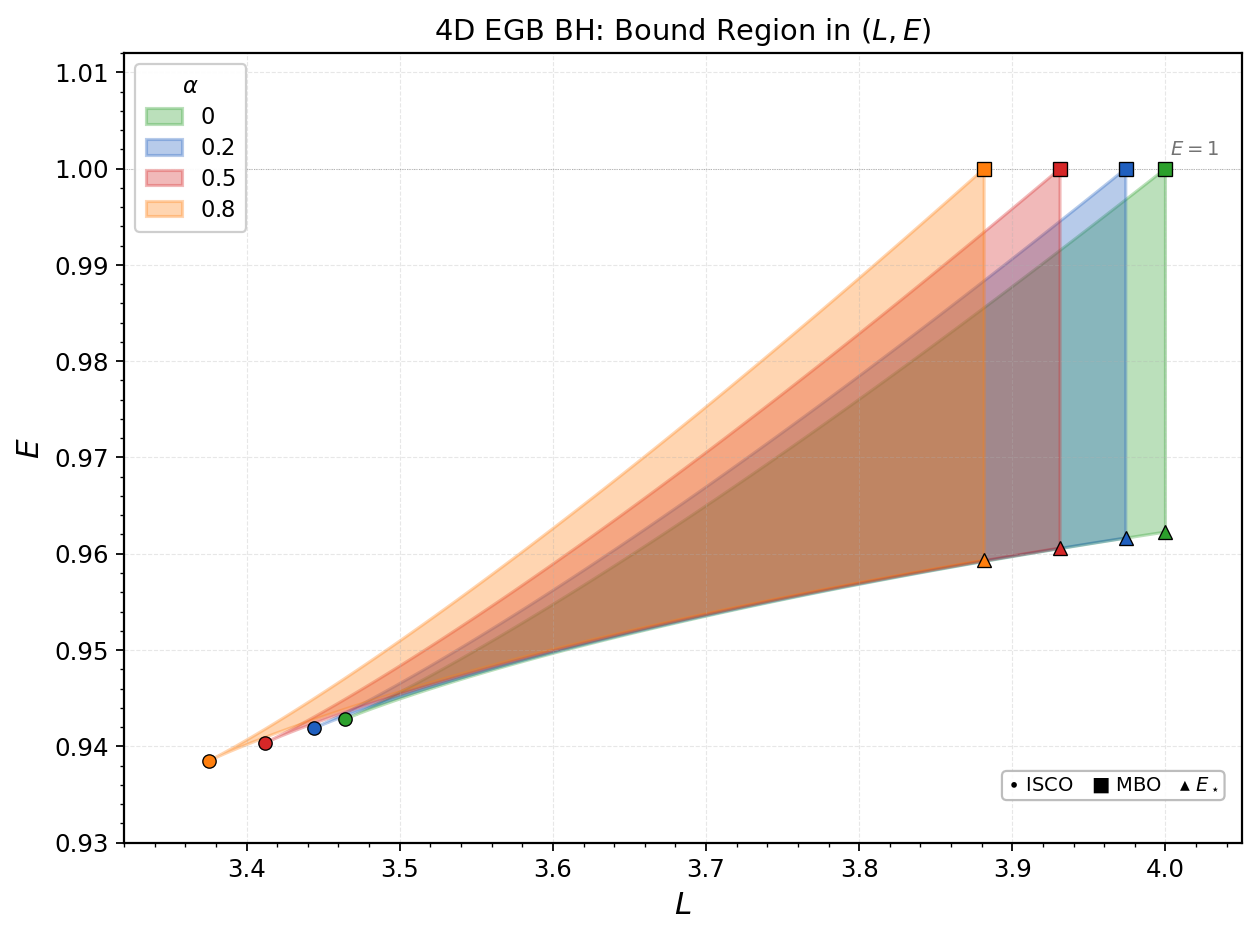}
\caption{Allowed $(L, E)$ parameter space for bound EMRIs. The permitted domains across different EGB couplings $\alpha$ are denoted by the shaded wedges. The unstable and stable circular orbits are represented by the upper and lower curved boundaries, respectively, which converge at the ISCO (circles). The right-side vertical boundary connects the MBO ($E=1$, squares) down to the corresponding stable circular orbit (triangles). By increasing $\alpha$, the entire parameter space is systematically shifted toward lower energies and angular momenta.}
\label{fig:2}
\end{figure}

\par
By numerically evaluating these boundaries, the allowed $(L, E)$ parameter space for bound EMRIs is mapped in Fig.~\ref{fig:2}. In the classical Schwarzschild limit ($\alpha = 0$), the standard analytical benchmarks are exactly recovered by our framework: the ISCO at $r_{\rm ISCO} = 6$ ($L_{\rm ISCO} = 2\sqrt{3} \approx 3.464$, $E_{\rm ISCO} = 2\sqrt{2}/3 \approx 0.943$), the MBO at $r_{\rm MBO} = 4$ ($L_{\rm MBO} = 4$), and the associated stable orbit at $r_\star = 12$ ($E_\star = \sqrt{25/27} \approx 0.962$). As the EGB coupling parameter $\alpha$ is increased, a systematic shift of the allowable parameter space toward lower angular momenta and energies is observed. Both the ISCO and MBO are dragged deeper into the strong-field regime by the higher-curvature corrections, whereby the requisite angular momentum for bound motion is reduced by approximately $2\%$ to $3\%$ as $\alpha \to 0.8$. A unique precessing trajectory is dictated by each coordinate within the shaded wedges of Fig.~\ref{fig:2}, which is systematically classified via the zoom-whirl taxonomy in the subsequent discussion.

\par
These bound periodic trajectories are classified using the rational taxonomy $(z, w, v)$ introduced in Ref.~\cite{Levin:2008mq}. Herein, the zoom number (radial leaves) is denoted by $z$, the whirl number per leaf is given by $w$, and the azimuthal phase shift between successive apastra is dictated by $v$. The fractional frequency ratio $q$ is defined by these integers as:
\begin{equation}
\label{eq-16}
q \equiv \frac{\omega_\phi}{\omega_r} - 1 = w + \frac{v}{z}.
\end{equation}
For a bound orbit with conserved constants $(L, E)$, the azimuthal angle accumulated during a single radial period is given by~\cite{Glampedakis:2002cb}
\begin{equation}
\label{eq-17}
\Delta\phi_r = 2 \int_{r_p}^{r_a} \frac{L}{r^2 \sqrt{E^2 - V_{\rm eff}(r)}} dr = 2\pi(1+q).
\end{equation}
To bypass the integrable singularities at the turning points ($r_{p,a}$), the integration domain is regularized via the trigonometric substitution
\begin{equation}
\label{eq-18}
r(u) = \frac{r_a + r_p}{2} + \frac{r_a - r_p}{2}\sin u, \quad u \in \left[-\frac{\pi}{2}, \frac{\pi}{2}\right].
\end{equation}
By this mapping, the inverse square-root divergence at the boundaries is explicitly canceled. Because $q(E)$ monotonically increases with energy for a fixed $L$ within the bound domain, Brent's method is employed to pinpoint the precise specific energy $E^*$ by which the target rational ratio $q$ is yielded~\cite{Glampedakis:2002cb}. When attempting to trace the spatial trajectory $r(\phi)$, it is found that direct numerical integration of the first-order geodesic equation $(dr/d\phi)^2 = (r^4/L^2)[E^2 - V_{\rm eff}(r)]$ is intrinsically unstable, a consequence of the alternating sign of $dr/d\phi$ at the turning points. This ambiguity is circumvented by differentiating this relation with respect to $\phi$, whereby a second-order ordinary differential equation is yielded
\begin{equation}
\label{eq-19}
\frac{d^2 r}{d\phi^2} = \frac{2 r^3}{L^2}\left[E^2 - V_{\rm eff}(r)\right] - \frac{r^4}{2 L^2}V_{\rm eff}'(r).
\end{equation}
This initial-value problem $\vec{y} = (r, dr/d\phi)^{\rm T}$ is subsequently integrated via a high-precision Runge-Kutta scheme, initialized at the apastron: $r(0) = r_a$ and $dr/d\phi|_{0} = 0$.

\begin{figure*}[htbp]
\centering
\includegraphics[width=12cm,height=6cm]{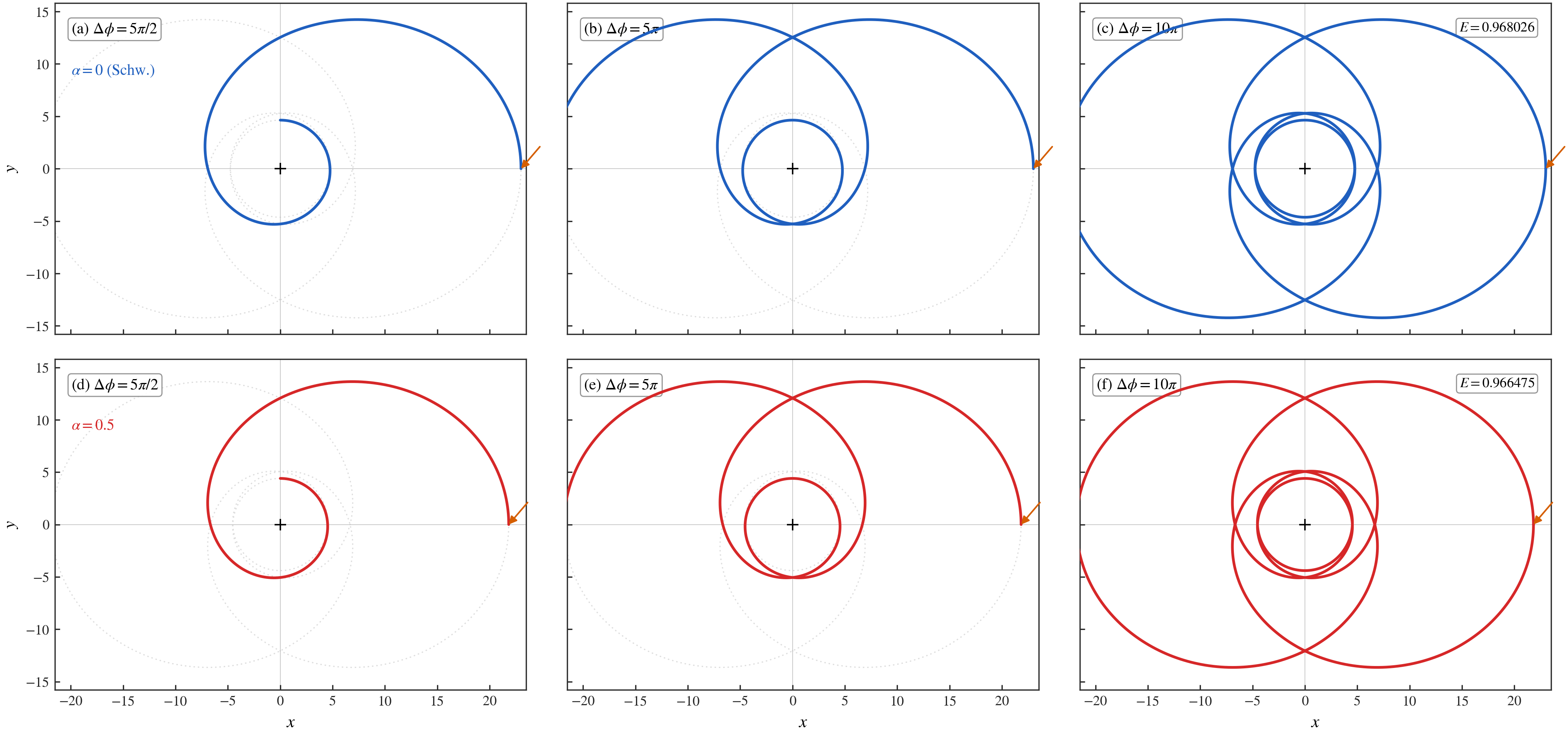}
\caption{Evolution of a $(2,1,1)$ periodic orbit for the Schwarzschild ($\alpha = 0$, top) and 4D EGB ($\alpha = 0.5$, bottom) black holes at a fixed $\varepsilon = 0.5$. The trajectory is captured by snapshots at accumulated angles $\Delta\phi = 5\pi/2$, $5\pi$, and $10\pi$ (closure). Traversed and pending paths are denoted by solid and dotted lines, respectively, while the initial apastron is marked by arrows. It is demonstrated that the periastron whirl is altered by strong-field EGB corrections, whereas the weak-field apastron is left visually unaffected.}
\label{fig:3}
\end{figure*}

\par
A $(2,1,1)$ periodic orbit ($q = 1.5$) is visualized in Fig.~\ref{fig:3}. Because $\Delta\phi_r = 5\pi$ is accumulated by a single radial leaf, a total azimuthal phase of $\Delta\phi = 10\pi$ is required for full orbital closure ($z=2$). By fixing $\varepsilon = 0.5$, $L \approx 3.732$ and $E^* \approx 0.9680$ are yielded by the root-finding procedure for the Schwarzschild baseline ($\alpha = 0$). For the 4D EGB black hole ($\alpha = 0.5$), the dynamically allowed space is shifted, resulting in $L \approx 3.672$ and $E^* \approx 0.9665$. As depicted in Fig.~\ref{fig:3}, the far-field apastron trajectories ($r \gtrsim 10$) are visually identical between the two theories. However, due to the rapid $\mathcal{O}(\alpha/r^4)$ spatial decay of the higher-curvature corrections, it is ensured that metric deviations are strictly localized near the periastron. The whirl frequency is consequently altered by this modified local potential well, whereby a distinct, measurable azimuthal phase shift is accumulated by the test particle in EGB gravity during the strong-field plunge.

\par
\begin{figure*}[htbp]
\centering
\includegraphics[width=14cm,height=5.5cm]{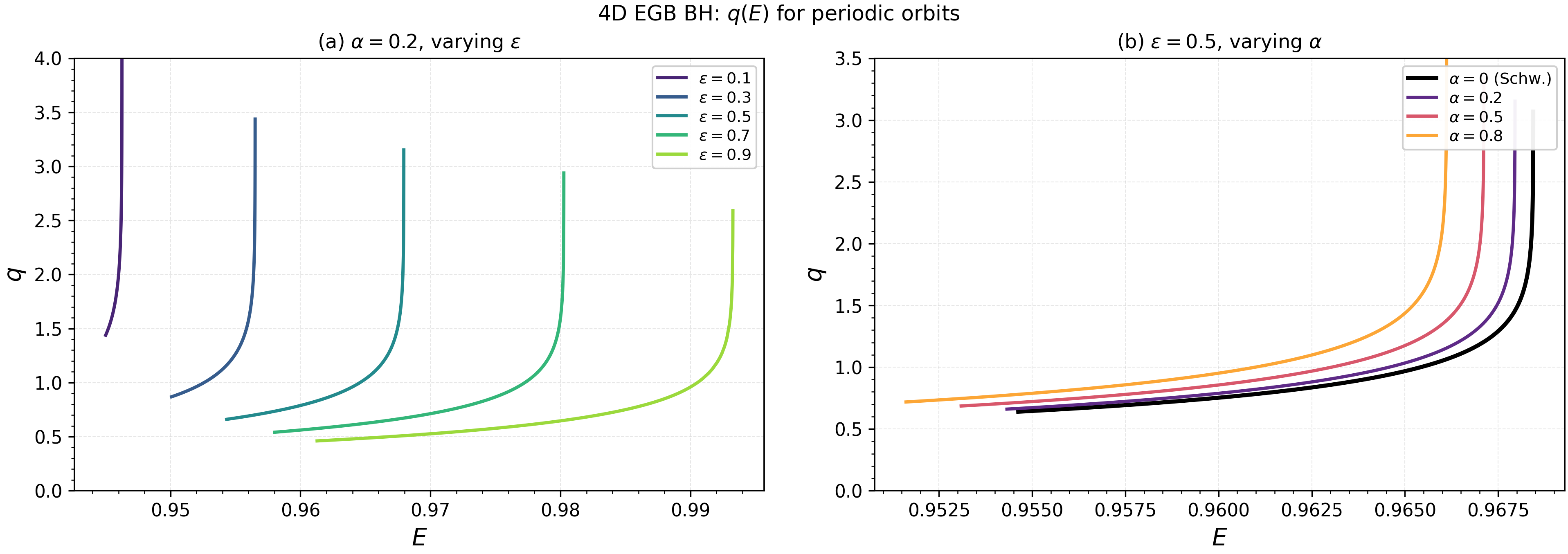}
\caption{Frequency ratio $q(E)$ for bound periodic EMRIs. (a) The effect of varying the fractional angular momentum $\varepsilon$ at a fixed EGB coupling $\alpha = 0.2$ is shown. (b) The effect of varying $\alpha$ at a fixed $\varepsilon = 0.5$ is shown, referencing the Schwarzschild limit ($\alpha=0$, black curve). Each profile is observed to span from the stable circular orbit limit to the unstable circular orbit asymptote.}
\label{fig:4}
\end{figure*}

To systematically map the accessible $(z, w, v)$ orbital topologies, the frequency ratio $q(E)$ is evaluated continuously across the dynamically allowed energy domain $E \in \left(\sqrt{V_{\rm min}}, \sqrt{V_{\rm max}}\right)$. The results are plotted in Fig.~\ref{fig:4}.

As illustrated in Fig.~\ref{fig:4}, $q(E)$ increases strictly monotonically with energy. In the low-energy limit, a convergence to a stable circular configuration is exhibited by the orbit, and $q$ is saturated to a finite minimum $q_{\rm min}$, which is characteristic of the small-amplitude epicyclic periastron advance. Conversely, as $E \to \sqrt{V_{\rm max}}^-$, a diverging proper time is spent by the test particle lingering near the unstable circular orbit. By this behavior, the strong-field whirling phase is indefinitely prolonged, whereby a vertical asymptote ($q \to \infty$) indicative of the homoclinic limit is produced.

\par
The impact of the fractional angular momentum $\varepsilon$ is isolated in Panel (a). For $L \to L_{\rm ISCO}$ ($\varepsilon \to 0$), the allowable energy is restricted to a narrow band by the shallow potential well. As $L \to L_{\rm MBO}$ ($\varepsilon \to 1$), this domain is broadened significantly toward the escape threshold ($E_{\rm max} \to 1$), while the concurrent decrease in $q_{\rm min}$ reflects a reduced periastron advance rate for the associated stable circular orbits. The imprints of the EGB higher-curvature corrections at a fixed $\varepsilon = 0.5$ are highlighted in Panel (b). In the classical Schwarzschild limit ($\alpha = 0$), the centrifugal barrier permits bound periodic motion up to $E_{\rm max} \approx 0.9684$. When the GB coupling is increased to $\alpha = 0.8$, this potential barrier is elevated and the threshold is marginally lowered to $E_{\rm max} \approx 0.9662$, by which the entire $q(E)$ curve is translated to the left. Notably, this global energetic displacement remains remarkably small---at the sub-percent level of $\sim 0.2\%$. This quantitative disparity precisely reflects the underlying geometry: the EGB metric deviation scales as $\mathcal{O}(\alpha/r^4)$. Although it is dictated by this rapid spatial decay that the corrections are sufficiently strong near the periastron to alter the local whirl phase (as visualized in Fig.~\ref{fig:3}), they are far too localized to substantially modify the global conservative energy budget of the orbit.

\par
Adopting a complementary perspective, the orbital energy is fixed at $E \in (E_{\rm ISCO}, 1)$, and the frequency ratio $q$ is evaluated as a continuous function of the specific angular momentum $L$. For a given energy, bound oscillatory motion is strictly confined to an admissible window $L \in [L_{\rm low}, L_{\rm high}]$. These boundaries correspond to the roots of $V_{\rm eff}(r; L) = E^2$ degenerating into circular orbits ($V_{\rm eff}' = 0$). The asymptotic behavior of $q(L)$ at these limits is governed by the local geometry of the effective potential. When $L \to L_{\rm high}^-$, the orbit is shrunken to the minimum of the potential well ($r_{\rm st}$, where $V_{\rm min} = E^2$). Through a small-perturbation expansion, a harmonic oscillator equation for the radial coordinate is yielded, whereby the frequency ratio is reduced to the finite epicyclic limit $q \to q_{\rm min} = 1/k - 1$. Here, the radial restoring force dictates $k^2 = r_{\rm st}^4 V_{\rm eff}''(r_{\rm st}) / (2L^2)$. Notably, as the orbital energy is reduced toward the ISCO limit ($E \downarrow E_{\rm ISCO}$), $V_{\rm eff}'' \to 0$ is given by the inflection point, causing $q_{\rm min} \to \infty$ as the admissible $L$ window is collapsed to a single degenerate point. When $L \to L_{\rm low}^+$, the peak of the potential barrier ($r_{\rm us}$) is approached by the periastron. When the potential is locally expanded as $V_{\rm eff}(r) \simeq V_{\rm max} - \frac{1}{2}\kappa^2(r-r_{\rm us})^2$, where $\kappa^2 = |V_{\rm eff}''(r_{\rm us})|$, it is found that the energy deficit scales linearly with the angular momentum variation: $V_{\rm max} - E^2 \propto (L - L_{\rm low})$. Consequently, a diverging proper time is spent by the test particle whirling near the barrier, and a logarithmic divergence is developed by the azimuthal accumulation in this homoclinic limit:
\begin{equation}
\label{eq-20}
q \simeq \frac{L}{\pi \kappa r_{\rm us}^2} \ln \left( \frac{1}{L - L_{\rm low}} \right) + \mathcal{O}(1),
\end{equation}
where $\kappa$ is the coordinate Lyapunov exponent of the unstable circular orbit. To numerically resolve this steep logarithmic gradient, a geometric mesh is employed so that sampling points can be densely clustered near $L_{\rm low}$.

\begin{figure*}[htbp]
\centering
\includegraphics[width=14cm,height=5.5cm]{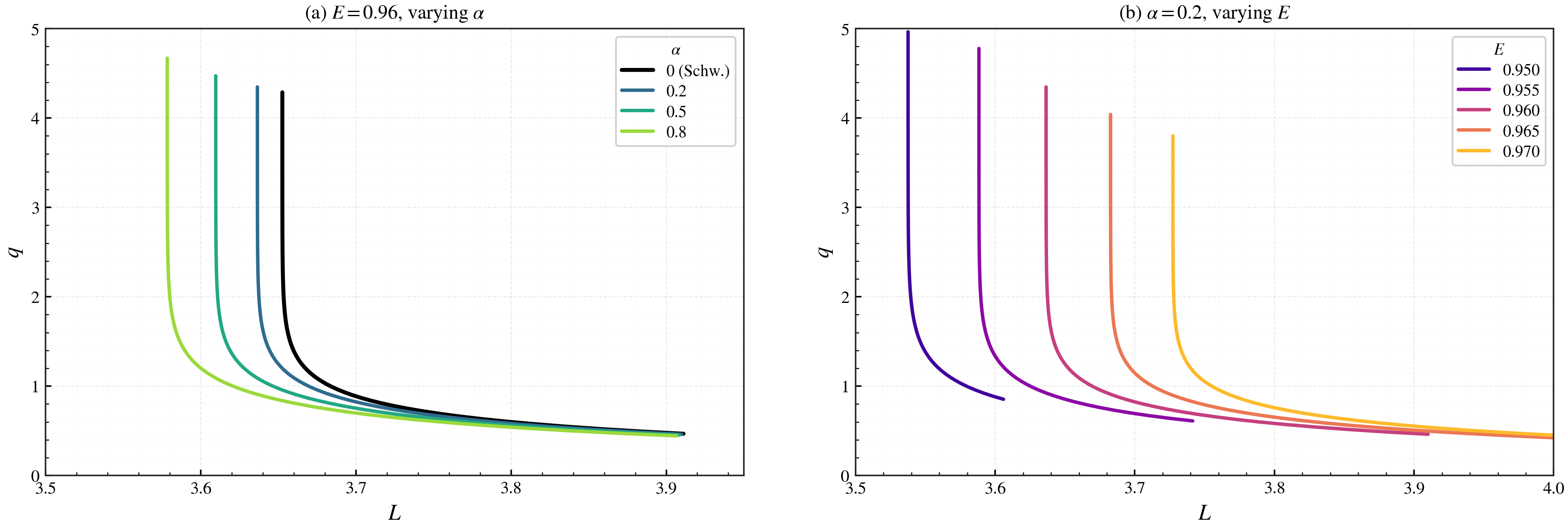}
\caption{Frequency ratio $q(L)$ for bound EMRIs. (a) The effect of varying the EGB coupling $\alpha$ at a fixed energy $E=0.96$ is illustrated. (b) The effect of varying $E$ at a fixed $\alpha = 0.2$ is shown. Each profile spans from the logarithmic homoclinic divergence at $L_{\rm low}$ (unstable circular orbit) to the finite epicyclic limit at $L_{\rm high}$ (stable circular orbit).}
\label{fig:5}
\end{figure*}

\par
The resulting $q(L)$ profiles are visualized in Fig.~\ref{fig:5}. The effect of the EGB coupling $\alpha$ is isolated in Panel (a). It is revealed by the weak-field expansion of the metric function, $f(r) \simeq 1 - 2/r + 2\alpha/r^4$, that the quantum correction acts as a short-range repulsive force. The centrifugal barrier is elevated by this repulsion, necessitating a smaller angular momentum to maintain the same binding energy. Thus, relative to the Schwarzschild baseline ($\alpha = 0$), the entire admissible $L$ window is systematically shifted to the left when $\alpha$ is increased. The energy dependence at a fixed $\alpha = 0.2$ is illustrated in Panel (b). As the energy approaches the marginally bound threshold ($E \to 1$), bound orbits with significantly larger angular momenta are accommodated by the potential well, by which the $[L_{\rm low}, L_{\rm high}]$ window is caused to broaden and shift right. Because distinct parameter-space translations are induced by $\alpha$ and $E$, the degeneracy in the mapping from an observed $(z,w,v)$ topological orbit to its physical parameters is broken. In EMRI observations, the strong-field whirls are predominantly modified by these localized short-range repulsions, thereby imprinting distinct, non-degenerate signatures onto the discrete frequency spectrum of the gravitational waveforms.

\par
To explicitly map these abstract frequency ratios into physical trajectories, a gallery of 12 periodic orbits in the 4D EGB spacetime is constructed, wherein $\alpha = 0.2$ and $\varepsilon = 0.5$ are fixed. Spatial closure in the equatorial plane requires that the total azimuthal accumulation after $z$ radial periods be an integer multiple of $2\pi$. When the rational classification $q = w + v/z$ is utilized, this closure condition is naturally dictated as~\cite{Levin:2008mq}
\begin{equation}
\label{eq-21}
\Phi_{\rm tot} = z \Delta\phi_r = 2\pi(z + zw + v).
\end{equation}
The total number of full $2\pi$ revolutions completed by the particle per closed orbit is defined by the integer $(z + zw + v)$. Herein, precise geometric meanings are carried by the topological numbers: the number of distinct apastron ``petals'' is dictated by $z$, governing the $z$-fold rotational symmetry; the angular skip between consecutive apastra is represented by $v$, effectively tracing out a $\{z/v\}$ star polygon; and the extra full whirls executed near the periastron are counted by $w$. It is ensured by the coprimality condition $\gcd(z,v)=1$ that the trajectory forms a single, continuously connected path without degenerating into sub-orbits. For each target topology $q^*$, the unique energy $E^*$ is numerically pinpointed, and the regularized second-order ODE \eqref{eq-19} is integrated from an initial apastron strictly up to $\phi = \Phi_{\rm tot}$, whereby spatial closure is verified by recovering the initial apastron radius.

\begin{figure*}[htbp]
\centering
\includegraphics[width=14cm,height=18cm]{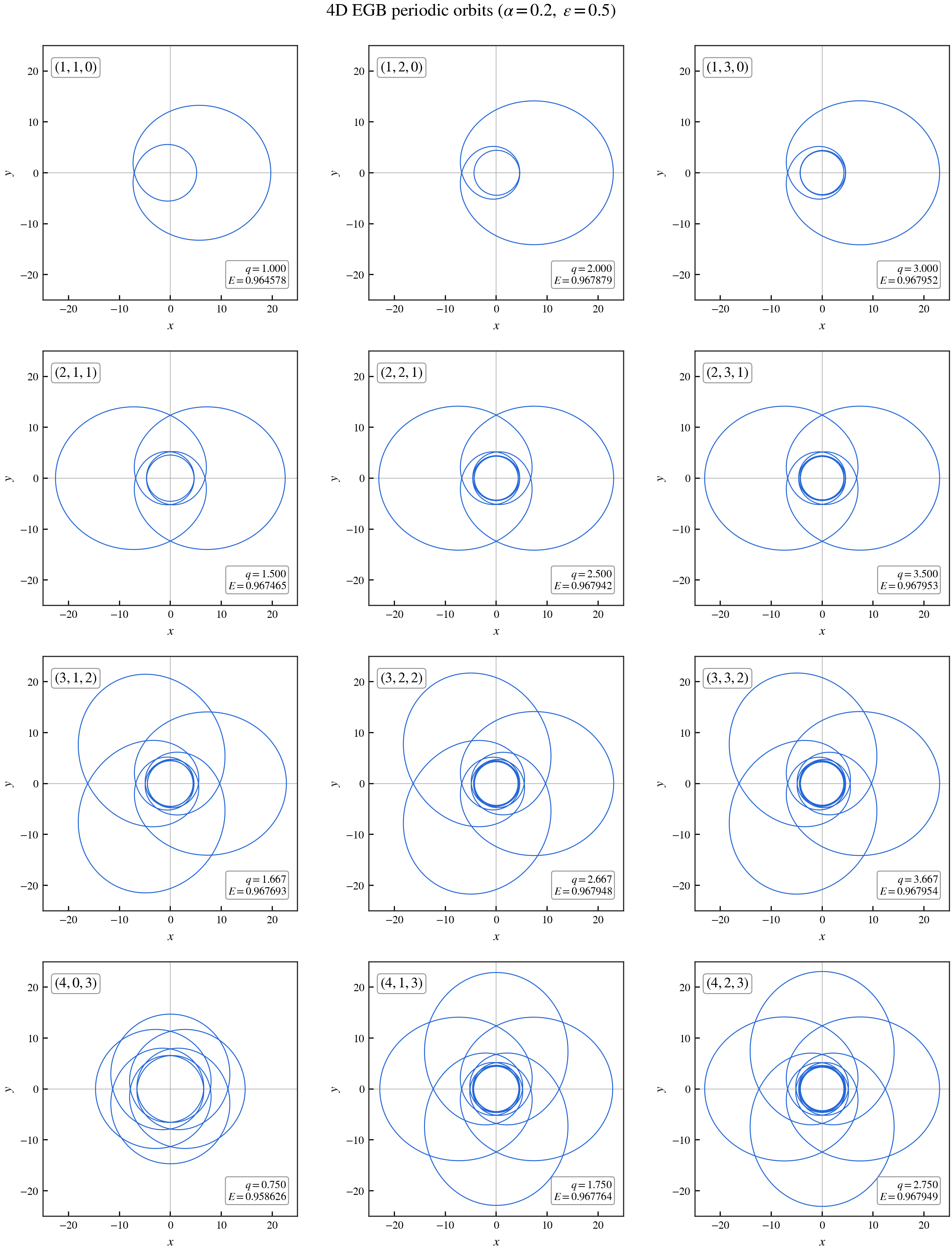}
\caption{A $4 \times 3$ gallery of $(z,w,v)$ periodic EMRIs in the 4D EGB spacetime ($\alpha = 0.2, \varepsilon = 0.5$). Increasing leaf numbers $z \in \{1,2,3,4\}$ are shown in the rows, with the vertex shift fixed at $v=z-1$ so that single-path closure ($\gcd(z,v)=1$) is ensured. Increasing periastron whirl numbers $w$ are shown in the columns. The target frequency ratio $q$ and required specific energy $E^*$ are annotated for each orbit.}
\label{fig:6}
\end{figure*}

\par
A visual atlas of EMRIs is provided by the resulting spatial topologies, which are systematically organized in Fig.~\ref{fig:6}. When moving down the columns (increasing $z$ at a fixed $w$), more apastron petals are developed by the trajectory in the weak-field region, while a similar strong-field winding character is maintained. Conversely, when moving across the rows (increasing $w$ at a fixed $z$), the petal count remains invariant, but the periastron dynamics are observed to grow profoundly intricate. Dynamically, larger orbital energies $E^*$ that approach the local peak of the effective potential are demanded by higher-$w$ topologies. In this homoclinic limit, the particle dives deeper into the strong-field regime, where an escalating number of dense $2\pi$ windings around the central black hole are executed before escaping. It is precisely during these localized, high-curvature whirl phases that the exposure of the particle to the $\mathcal{O}(\alpha/r^4)$ EGB metric deviations is maximized. For EMRIs targeted by future space-borne interferometers (e.g., LISA, Taiji), prolonged, high-frequency gravitational-wave bursts are generated by these strong-field zoom-whirl cycles. Because the energy $E^*$ required to sustain a specific $w$-whirl topology is measurably altered by the short-range EGB repulsion, an anomalous secular phase shift will be imprinted onto these waveform bursts.

\section{Gravitational-wave emission from periodic orbits and multi-messenger observables}
\label{sec:3}
\par
\begin{figure*}[htbp]
\centering
\includegraphics[width=14cm,height=6cm]{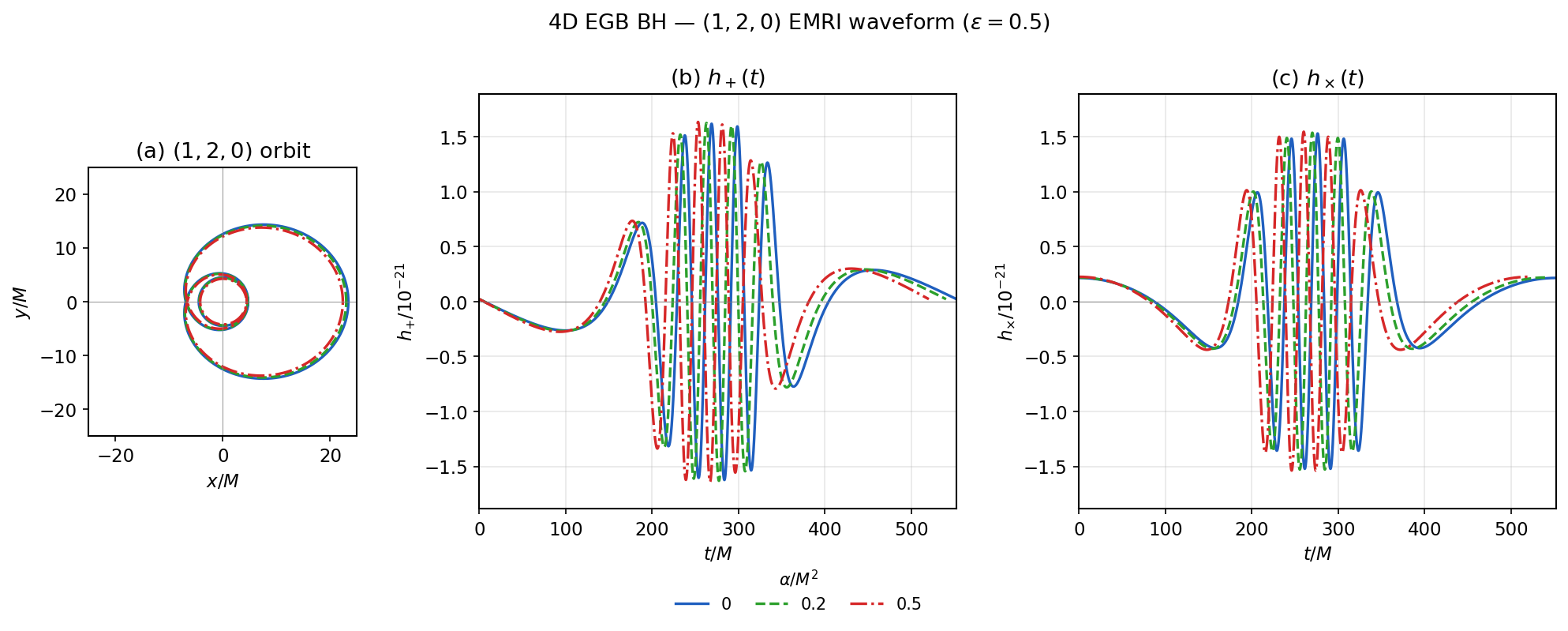}
\caption{Gravitational-wave emission from the $(1,2,0)$ periodic EMRI ($q=2, \varepsilon=0.5$). (a) Spatial trajectory in the equatorial plane for $\alpha \in \{0, 0.2, 0.5\} M^2$. (b) Plus polarization $h_+$. (c) Cross polarization $h_\times$. The observer is located at $(\Theta, \Phi) = (\pi/4, \pi/4)$ with a mass ratio $m/M=10^{-5}$ at $D_L=200$ Mpc. The time-domain compression is visually apparent: the orbital period is shortened by the repulsive EGB core, whereby the high-frequency periastron bursts are caused to arrive progressively earlier for larger $\alpha$.}
\label{fig:7}
\end{figure*}

With the exact topological periodic orbits in the 4D EGB spacetime having been established, these strong-field trajectories are subsequently translated into direct observational signatures. EMRIs, wherein a stellar-mass compact object orbits a supermassive black hole, are characterized by the generation of thousands of continuous gravitational-wave cycles~\cite{LISA:2017pwj,Hu:2017mde,TianQin:2015yph}. To compute the theoretical waveform, the orbital kinematics must be mapped from the azimuthal parameter $\phi$ to the coordinate time $t$ of a distant observer. By utilizing the conserved quantities $E = f(r)\dot{t}$ and $L = r^2 \dot{\phi}$, the coordinate time is integrated alongside the spatial trajectory via~\cite{Barack:2003fp}
\begin{equation}
\label{eq-22}
\frac{dt}{d\phi} = \frac{E r^2}{L f(r)}.
\end{equation}
By numerically integrating the coupled system $\{r(\phi), t(\phi)\}$, the parametric evolution is obtained, which is subsequently re-sampled onto a uniform coordinate time grid using cubic spline interpolation so that the Cartesian kinematics $x^i(t)$ and $v^i(t)$ can be extracted~\cite{Babak:2006uv}. The gravitational radiation in the weak-field approximation is evaluated utilizing the semi-relativistic quadrupole formalism. For a test mass $m$, the trace-reduced spatial quadrupole tensor $I^{ij} = m x^i x^j$ yields the strain tensor $h^{ij} = (2G/c^4 D_L) \ddot{I}^{ij}$~\cite{Gair:2005ih}. When the second time derivative is expanded, the following expression is obtained
\begin{equation}
\label{eq-23}
h^{ij} = \frac{2m}{D_L} \left( a^i x^j + a^j x^i + 2 v^i v^j \right),
\end{equation}
where $D_L$ represents the luminosity distance. Because the motion is strictly confined to the equatorial plane ($z=0$), all orthogonal kinematic components vanish ($v_z = a_z = 0$), leaving $h^{xx}$, $h^{yy}$, and $h^{xy}$ as the only non-zero components. To extract the physical observables, the spatial strain must be projected into the transverse-traceless (TT) gauge. For a distant observer located at a viewing angle $(\Theta, \Phi)$, the line-of-sight vector is defined as $\hat{n} = (\sin\Theta\cos\Phi, \sin\Theta\sin\Phi, \cos\Theta)$. The transverse plane is spanned by the local orthogonal polarization basis vectors~\cite{Babak:2006uv}
\begin{align}
\label{eq-24}
\hat{e}_\Theta &= (\cos\Theta\cos\Phi, \cos\Theta\sin\Phi, -\sin\Theta), \nonumber \\
\hat{e}_\Phi &= (-\sin\Phi, \cos\Phi, 0).
\end{align}
The two independent gravitational-wave polarizations, $h_+$ and $h_\times$, are then strictly constructed by projecting $h^{ij}$ onto this TT basis
\begin{equation}
\label{eq-25}
h_+ = \frac{1}{2} \left( \hat{e}_\Theta^i \hat{e}_\Theta^j - \hat{e}_\Phi^i \hat{e}_\Phi^j \right) h_{ij}, \quad h_\times = \hat{e}_\Theta^i \hat{e}_\Phi^j h_{ij}.
\end{equation}
A representative EMRI system is simulated by utilizing the $(z,w,v) = (1, 2, 0)$ periodic precessing orbit ($q = 2, \varepsilon = 0.5$). The characteristic mass ratio is set to $m/M = 10^{-5}$, and the observer is positioned at $(\Theta, \Phi) = (\pi/4, \pi/4)$ with a luminosity distance of $D_L = 200~{\rm Mpc}$. The resulting waveforms for the GB couplings $\alpha \in \{0, 0.2, 0.5\} M^2$ are presented in Fig.~\ref{fig:7}.

\par
In Fig.~\ref{fig:7}, the classic EMRI ``zoom-whirl'' structure is distinctively displayed: a low-frequency, low-amplitude envelope generated during the extended apastron zoom phase is sharply punctuated by dense bursts of high-frequency radiation emitted during the intense periastron whirls. Crucially, although the $(1,2,0)$ spatial topology is visually identical across the three cases, the temporal evolution of the radiation is rendered exquisitely sensitive to the underlying gravity theory. As has been demonstrated analytically, the effective potential barrier is elevated by the $\mathcal{O}(\alpha/r^4)$ repulsive core. Consequently, it is required that a progressively lower specific energy and angular momentum be possessed to maintain the exact $q=2$ topological resonance as $\alpha$ is increased. By this energetic shift, the coordinate orbital period $T$ is tightly compressed from $552 M$ in the Schwarzschild baseline ($\alpha=0$) down to $540 M$ ($\alpha=0.2$) and $520 M$ ($\alpha=0.5$). In the time domain, a systematic temporal compression of the waveform is directly yielded by this effect. When initialized from an identical phase, the high-frequency periastron bursts are observed to arrive significantly earlier for larger $\alpha$ values. Because a physical EMRI signal is integrated over $\sim \mathcal{O}(M/m)$ cycles before the plunge, a massive macroscopic phase shift will be rapidly accumulated from this secular, microscopic period compression. By such severe cumulative dephasing, it is confirmed that a robust, non-degenerate observational channel to constrain 4D EGB quantum corrections is provided by zoom-whirl dynamics.

\par
Although the temporal compression of the waveforms is visually evident, it must be emphasized that precise gravitational-wave data analysis is fundamentally predicated upon matched filtering, a technique which is exquisitely sensitive to the accumulated signal phase. For equatorial EMRIs, the radiation is dominated by the $l=m=2$ quadrupole mode. Because the spatial quadrupole tensor scales azimuthally as $\propto e^{i 2 \phi_{\rm orb}}$, the gravitational-wave phase is driven to evolve at exactly twice the orbital angular frequency: $\phi_{GW}(t) = 2\phi_{\rm orb}(t)$. This effect is quantified by defining the cumulative GW dephasing of a given EGB orbit relative to the classical Schwarzschild ($\alpha=0$) baseline as~\cite{Maggiore:2007}
\begin{equation}
\label{eq-26}
\Delta\phi_{GW}(t) = 2\left[ \phi^{(\alpha)}_{\rm orb}(t) - \phi^{(0)}_{\rm orb}(t) \right].
\end{equation}
All evaluated orbits are identically initialized at the apastron ($\phi^{(\alpha)}_{\rm orb}(0) = 0$). To ensure a mathematically rigorous comparison, the dephasing is strictly evaluated within the shared temporal window $t \in [0, \min_\alpha T^{(\alpha)}]$.

\begin{figure*}[htbp]
\centering
\includegraphics[width=14cm,height=12cm]{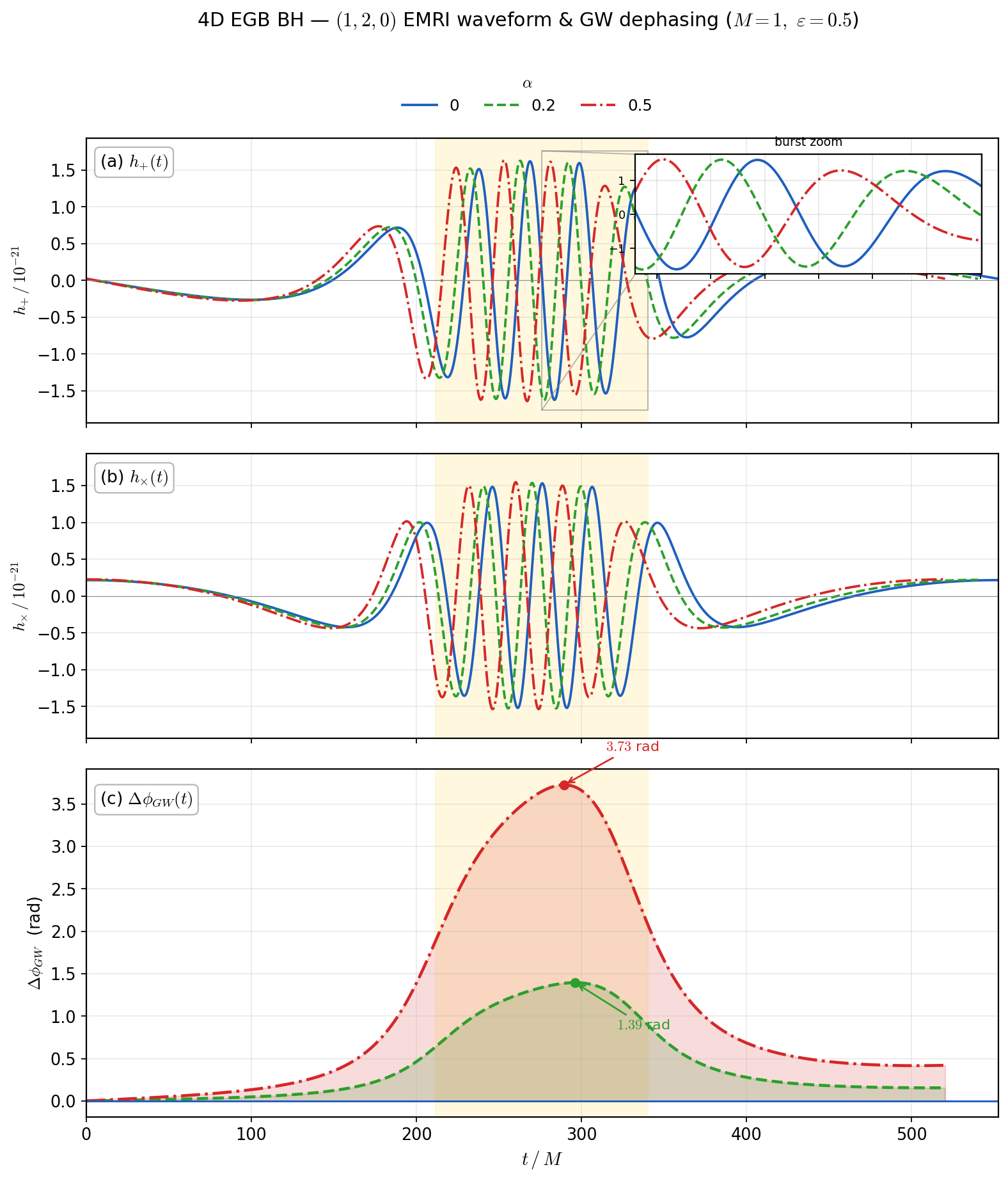}
\caption{Quantitative dephasing of the $(1,2,0)$ EMRI waveforms ($\varepsilon = 0.5$) for $\alpha \in \{0, 0.2, 0.5\} M^2$. Panels (a) and (b) overlay the $h_+$ and $h_\times$ polarizations, with the inset in (a) magnifying the progressive phase drift during periastron bursts (yellow band). In Panel (c), the explicit evolution of the cumulative GW dephasing $\Delta\phi_{GW}(t)$ relative to the $\alpha=0$ baseline is traced, revealing macroscopic peaks driven by the strong-field EGB modifications.}
\label{fig:8}
\end{figure*}

\par
In Fig.~\ref{fig:8}, a high-resolution overlay of the $(1,2,0)$ waveforms is presented alongside the explicit temporal evolution of $\Delta\phi_{GW}(t)$. In the inset of Fig.~\ref{fig:8}(a), the severe phase drift experienced as the particle traverses the strong-field periastron window (yellow shaded band) is clearly captured. The physical origin of this transient dephasing is cleanly isolated in Fig.~\ref{fig:8}(c). By rigorous topological definition, the identical $(1,2,0)$ class is shared by all three trajectories, whereby exactly $\phi_{\rm orb}(T) = 2\pi(1+q) = 6\pi$ per radial period must be accumulated. However, distinctive differences are exhibited by their time parameterizations. The orbital dynamics are globally accelerated by the $\mathcal{O}(\alpha/r^4)$ short-range repulsion, whereby the test particle in the EGB spacetime is caused to enter and complete its intense whirl phase strictly earlier than its Schwarzschild counterpart. By this kinematic discrepancy, $\Delta\phi_{GW}(t)$ is forced to surge rapidly during the periastron passage, reaching resolvable macroscopic peaks of $|\Delta\phi_{GW}|_{\max} \approx 1.39~{\rm rad}$ ($\alpha = 0.2$) and $3.73~{\rm rad}$ ($\alpha = 0.5$). The numerical integrity of these $\mathcal{O}(1)$ radian dephasings is particularly emphasized. When orbits are chosen too close to the marginally bound separatrix ($E \to 1$), extreme logarithmic winding (homoclinic divergence) is suffered, by which ODE integrations are inherently corrupted and the monotonic scaling of the orbital period is obscured. By securely anchoring the evaluations at $\varepsilon = 0.5$, these numerical artifacts are decisively evaded: the radial closure residuals are strictly returned to the initial apastron within $\lesssim 10^{-11}$, and the phase closure is precisely matched to the $6\pi$ analytical requirement to within $\mathcal{O}(10^{-5})$. Given that $\mathcal{O}(10^4 - 10^5)$ such orbits are completed over the mission lifetime of a typical EMRI system observed by space-borne interferometers, it is inevitable that a highly detectable macroscopic dephasing will be accumulated from this robust microscopic phase drift.

\par
While an intuitive picture of orbital deviations is provided by the time-domain dephasing, space-borne gravitational-wave observatories are fundamentally reliant upon frequency-domain matched filtering~\cite{Hughes:2005qb}. To bridge the dynamical results with direct observables, a precise spectral analysis of the emitted radiation is performed. Physically, exactly $\Phi_{\rm tot} = 6\pi$ per radial period $T$ is accumulated by the Levin-Perez-Giz $(1,2,0)$ periodic orbit. Because the azimuthal angle is advanced by an exact integer multiple of $2\pi$, the spatial kinematics, and consequently the quadrupole radiation strain, are returned precisely to their initial states. By this strict closure, perfect periodicity of the waveforms is guaranteed~\cite{Sundararajan:2007jg}
\begin{equation}
\label{eq-27}
h_{+,\times}(t+T) = h_{+,\times}(t).
\end{equation}
Mathematically, the expansion $h(t) = \sum_{k} c_k e^{2\pi i k t / T}$ is admitted by the Fourier series of such a strictly periodic signal. The continuous spectrum is thus strictly reduced to a discrete frequency comb, wherein power is entirely concentrated at harmonics $f_k = k/T$. These harmonic coefficients are numerically extracted via a Discrete Fourier Transform (DFT) uniformly sampled over a single period
\begin{equation}
\label{eq-28}
\tilde{h}^{(k)}_{+,\times} = \Delta t \sum_{n=0}^{N-1} h_{+,\times}(t_n) e^{-2\pi i k n / N},
\end{equation}
where $t_n = n\Delta t$ and $\Delta t = T/N$. Fifty sampling points per period ($N=50$) are utilized. The numerical robustness of this specific resolution has been rigorously validated: the aliased power above the Nyquist frequency ($f_{\rm Nyq} = N/2T$) remains strictly below $0.01\%$, the low-harmonic amplitudes deviate by $<1\%$ compared to an oversampled $N=4096$ reference, and time-frequency energy conservation (Parseval's theorem) is preserved to machine precision. To contextualize these spectra within the observational window of space-based interferometers, physical units are restored by assuming a central supermassive black hole of $M = 10^6 M_\odot$. The fundamental time scale is thus $1 M = G M / c^3 \approx 4.93~{\rm s}$. For a typical orbital period of $T \sim 550 M$, the entire harmonic comb is conveniently anchored squarely within the optimal mHz detection band by the fundamental frequency $f_1 = 1/T \approx 0.37~{\rm mHz}$.

\begin{figure*}[htbp]
\centering
\includegraphics[width=14cm,height=12cm]{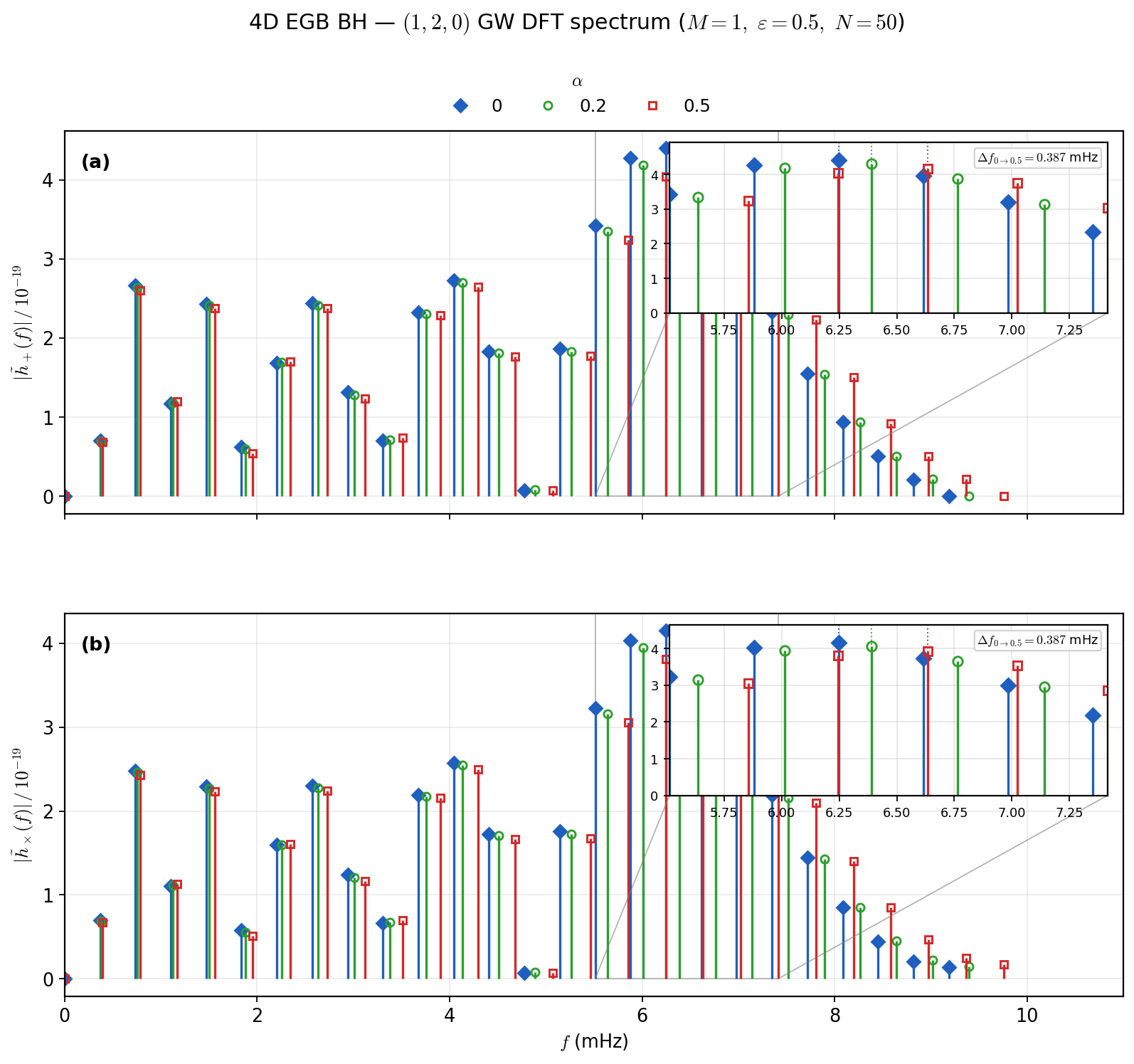}
\caption{Discrete frequency combs of the $h_+$ (a) and $h_\times$ (b) periodic waveforms ($M = 10^6 M_\odot, \varepsilon = 0.5$). Spectra for $\alpha = 0$, $0.2$, and $0.5$ are represented by blue diamonds, green circles, and red squares, respectively. The insets magnify the dominant spectral peak ($k \approx 17$) associated with the intense periastron burst. By the short-range EGB repulsion, the fundamental frequency $f_1$ is increased, whereby the comb is linearly stretched at higher harmonics and a highly resolvable macroscopic peak displacement is yielded.}
\label{fig:9}
\end{figure*}

The resulting harmonic spectra $|\tilde{h}^{(k)}_{+,\times}|$ for $\alpha \in \{0, 0.2, 0.5\} M^2$ are overlaid in Fig.~\ref{fig:9}. Although the spectra nearly coincide at lower harmonics, the combs are observed to fan out significantly as the harmonic index $k$ increases. The exact frequency-domain manifestation of the previously observed temporal compression is represented by this progressive spectral separation. Because the orbital period $T(\alpha)$ is shortened by the $\mathcal{O}(\alpha/r^4)$ EGB coupling, the fundamental frequency $f_1(\alpha)$ is inherently raised. Consequently, the absolute frequency shift of the $k$-th harmonic, i.e., $\Delta f_k = k \Delta f_1$, scales linearly with the harmonic number. The high-frequency burst emitted during the strong-field periastron whirl is dominated by these higher harmonics. As highlighted in the insets of Fig.~\ref{fig:9}, the dominant power peak is distinctively caused by this linear amplification to migrate from $6.25~{\rm mHz}$ in the classical Schwarzschild limit to $6.39~{\rm mHz}$ ($\alpha = 0.2$) and $6.63~{\rm mHz}$ ($\alpha = 0.5$). Since this uniform ``stretching'' of the frequency comb arises as a direct kinematic consequence of the modified strong-field spacetime geometry, a remarkably clean observational signature is offered. By the macroscopic shift of high-$k$ harmonic peaks, a reliable, amplitude-independent metric is provided, through which the short-range quantum corrections of the 4D EGB theory can be constrained via future EMRI detections.

\par
In the preceding analysis, the orbital topology was locked while the constants of motion $(E, L)$ were adjusted. An alternative, physically illuminating perspective emerges when $(E, L)$ are fixed and the manner in which the underlying gravity theory dictates the allowed topology is investigated. The orbital topology is governed by the rational frequency ratio $q = \Delta\phi_r / (2\pi) - 1$, where the azimuthal phase accumulation per radial period is given by
\begin{equation}
\label{eq-29}
\Delta\phi_r = 2 \int_{r_p}^{r_a} \frac{L}{r^2 \sqrt{E^2 - V_{\rm eff}(r; L, \alpha)}} dr.
\end{equation}

In the 4D EGB spacetime, the potential barrier maximum $V_{\max}$ is effectively elevated by the $+2\alpha/r^4$ repulsive core. For a fixed $(E, L)$, the periastron $r_p$ is pushed outward by a larger $\alpha$, whereby the particle is prevented from probing the deep strong-field regime. The extreme periastron whirls are dynamically ``unwound'' by this effect, rendering $q(\alpha; E, L)$ a strictly monotonically decreasing function. By exploiting this monotonicity, the mapping can be inverted: for any rational target topology $q^\star = w + v/z$, a unique coupling $\alpha^\star$ exists by which $q(\alpha^\star) = q^\star$ is satisfied. To demonstrate this topology selection cleanly, the constants are fixed at $E = 0.96$ and $L = 3.6534$. In the Schwarzschild limit ($\alpha = 0$), the dense, near-homoclinic $(1,2,0)$ baseline orbit ($q=2$) is yielded by these parameters. By numerically inverting Eq.~\eqref{eq-29}, the exact EGB couplings that organically select simpler topologies under the identical $(E, L)$ are pinpointed: the $(3,1,2)$ orbit ($q \approx 1.667$) is selected at $\alpha \approx 0.025~M^2$, the $(2,1,1)$ orbit ($q = 1.5$) at $\alpha \approx 0.055~M^2$, and the $(1,1,0)$ orbit ($q=1$) at $\alpha \approx 0.413~M^2$.

\begin{figure*}[htbp]
\centering
\includegraphics[width=\textwidth]{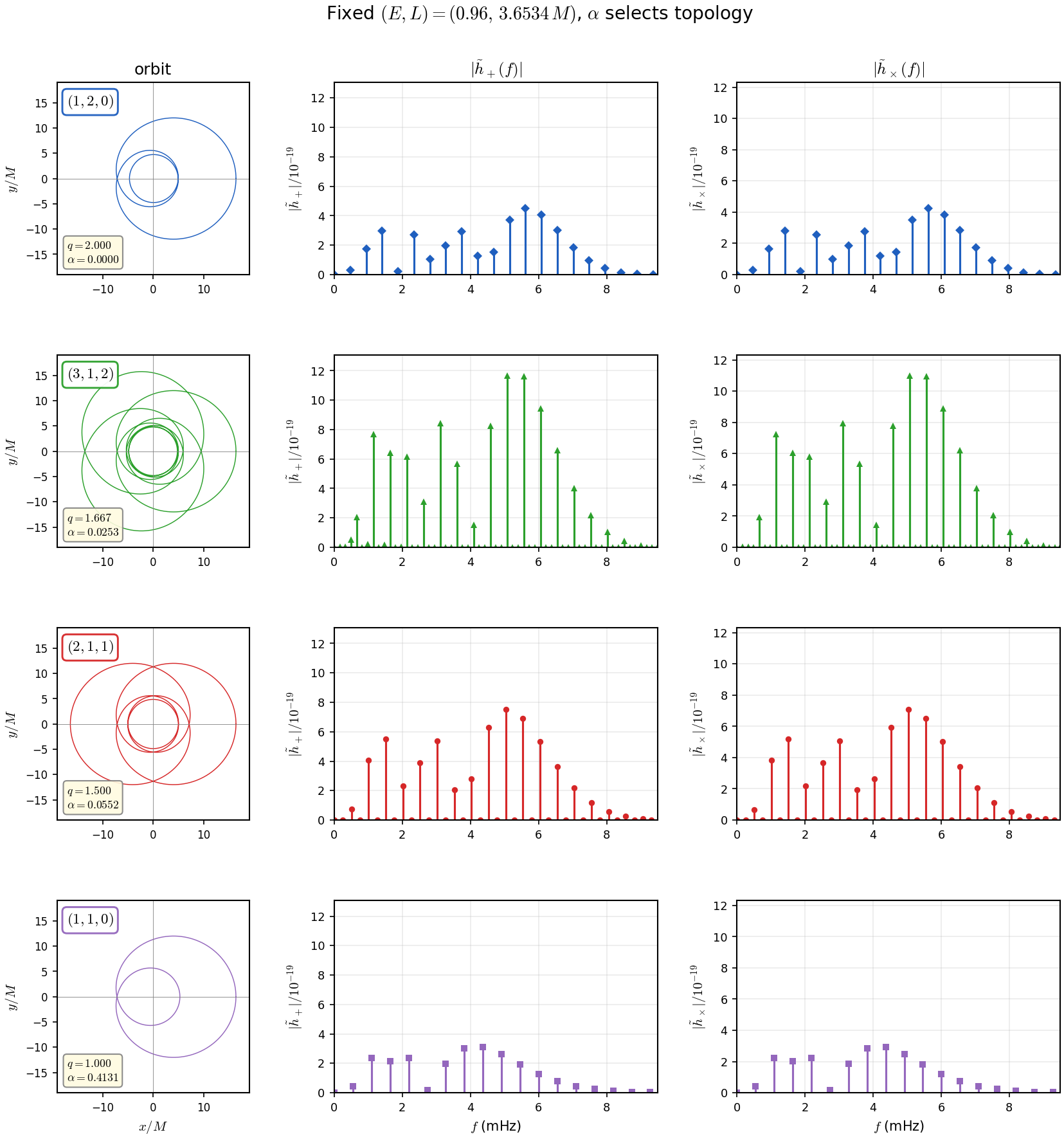}
\caption{Orbital morphologies and GW spectra selected by varying the GB coupling $\alpha$ under fixed constants of motion $(E = 0.96, L = 3.6534)$. Left: Spatial trajectories. Due to negligible far-field GB corrections, the same apastron ($r_a \approx 16.39$) is identically shared by all topologies; the macroscopic differences are strictly strong-field unwinding effects. Center/Right: Harmonic spectra for $|h_+|$ and $|h_\times|$. Longer periods $T$ are inherently possessed by more complex topologies (larger leaf integer $z$), whereby a denser frequency comb ($f_k = k/T$) is produced.}
\label{fig:10}
\end{figure*}

\par
The selected spatial morphologies and their corresponding DFT spectra are presented in Fig.~\ref{fig:10}. Because a rapid decay is exhibited by the $\mathcal{O}(\alpha/r^4)$ deviation, the outer turning point defined by $E^2 = V_{\rm eff}(r_a)$ remains virtually unperturbed. Consequently, an identical apastron distance is shared by all four disparate topologies. The pristine control group by which the strong-field geometric signature of the 4D EGB theory is isolated is served by the transition from a dense $(1,2,0)$ rosette to a simple $(1,1,0)$ ellipse. The spectral manifestations are similarly distinctive. Because $z$ radial oscillations are required for a $z$-leaf periodic orbit to close, its total coordinate period $T$ is rendered proportional to $z$. In the frequency domain, a smaller fundamental frequency $f_1 = 1/T$ is directly translated from a more complex topology (larger $z$), whereby a denser harmonic comb is yielded. Thus, a dense forest of spectral lines is exhibited by the $(1,2,0)$ and $(3,1,2)$ orbits, whereas a sparse, widely separated spectrum is displayed by the mathematically simpler $(1,1,0)$ topology. It should be noted that extracting physically reliable spectra from near-homoclinic orbits is notoriously difficult, as severe logarithmic phase divergence is suffered by standard ODE integrators. To guarantee the absolute integrity of these spectra, ultra-stringent tolerances ($\text{rtol} = 10^{-12}$, $\text{atol} = 10^{-14}$) were enforced, by which the radial closure residuals were firmly suppressed to $\sim \mathcal{O}(10^{-8})$. Furthermore, to preclude high-frequency aliasing for large-$z$ orbits, the DFT sampling size was dynamically scaled as $N = \max(64, 48z)$. By these rigorous treatments, it is ensured that the topological spectra presented herein act as exact and robust observables.

\par
To assess the detectability of these 4D EGB EMRI signals, their characteristic strains are evaluated against realistic noise models of upcoming space-based interferometers. In optimal matched filtering, the signal-to-noise ratio (SNR) squared is evaluated as
\begin{equation}
\label{eq-30}
\mathrm{SNR}^2 = 4 \int_0^\infty \frac{|\tilde{h}(f)|^2}{S_n(f)} df = \int_0^\infty \left[ \frac{h_c(f)}{h_n(f)} \right]^2 d\ln f,
\end{equation}
where $S_n(f)$ is the one-sided power spectral density of the detector. By changing the integration measure to $d\ln f$, the dimensionless characteristic strains for the dual-polarization signal and the detector noise are naturally defined as
\begin{equation}
\label{eq-31}
h_c(f) = 2f \sqrt{|\tilde{h}_+(f)|^2 + |\tilde{h}_\times(f)|^2}, \quad h_n(f) = \sqrt{f S_n(f)}.
\end{equation}
In a log-log plot, the accumulated $\mathrm{SNR}^2$ is directly represented by the visual area enclosed between $h_c$ and $h_n$. Because actual space missions operate over a finite lifetime $T_{\rm obs}$, the strictly periodic orbit must be truncated via a rectangular time window (here simulated as $T_{\rm obs} = 23.4\, T_{\rm orb}$), by which the discrete frequency comb is mathematically convolved with a sinc function. Signal power is leaked into adjacent bins by this spectral broadening, whereby a characteristic ``sinc-carpet'' sidelobe structure is formed. To resolve these highly oscillatory sidelobes without aliasing, a periodised Fast Fourier Transform is implemented over an ultra-dense temporal grid ($N=32768$ points, yielding $\approx 1400$ samples per period).

\begin{figure*}[htbp]
\centering
\includegraphics[width=\textwidth]{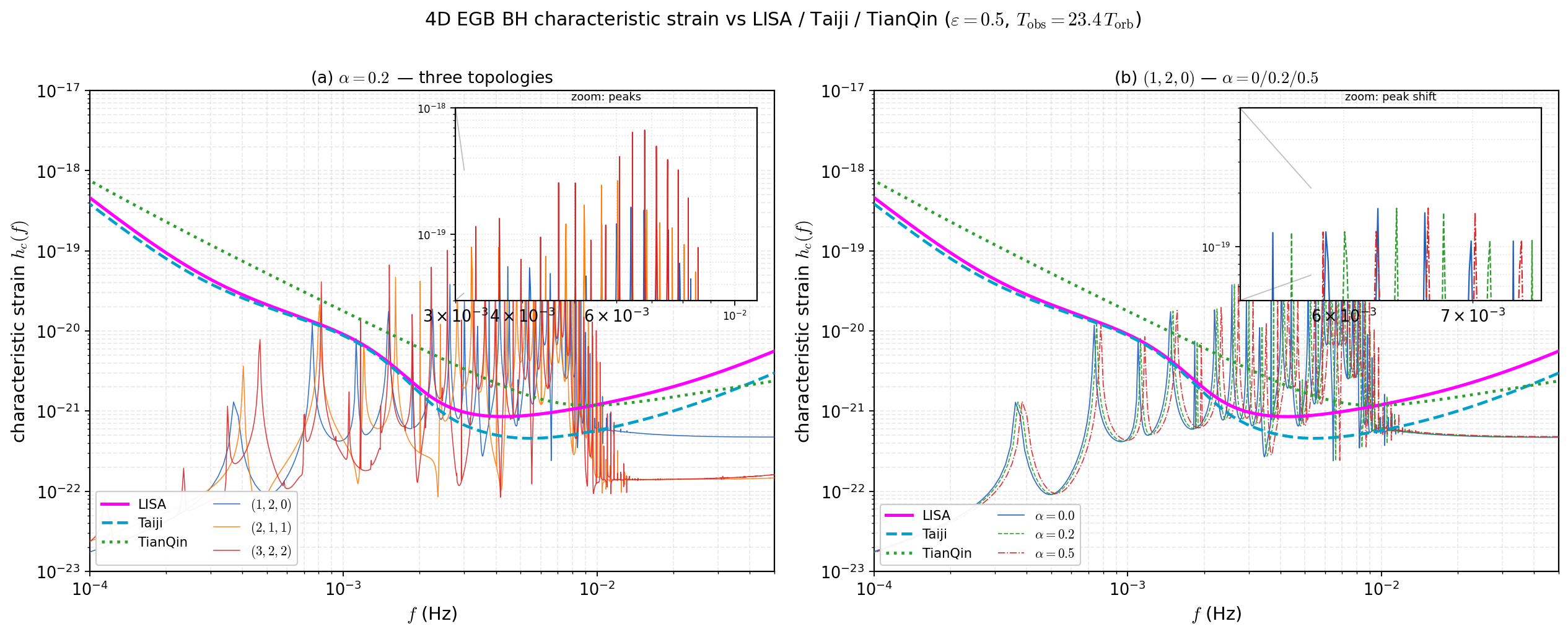}
\caption{Characteristic strains $h_c(f)$ of 4D EGB EMRI signals ($M = 10^6 M_\odot, \mu = 10 M_\odot, D_L = 200~{\rm Mpc}$) compared with the sensitivity curves $h_n(f)$ of LISA, Taiji, and TianQin. (a) Topological effect at fixed $\alpha = 0.2$. Prolonged strong-field bursts are radiated by complex orbits like $(3,2,2)$, yielding denser spectra and significantly higher SNRs. (b) EGB parameter effect for a fixed $(1,2,0)$ topology. Increasing $\alpha$ rigorously shifts the dominant periastron peaks to higher frequencies without amplitude degradation, offering a clean, high-SNR observational signature.}
\label{fig:11}
\end{figure*}

\par
In Fig.~\ref{fig:11}, the computed $h_c(f)$ is overlaid against specific instrumental sensitivities, assuming a prototypical system ($M = 10^6 M_\odot$, $\mu = 10 M_\odot$, $D_L = 200~{\rm Mpc}$). By the extreme mass-ratio, the high-frequency periastron bursts are naturally scaled squarely into the $2$--$10~{\rm mHz}$ interferometry window. Specifically, the optimal minimums of the detector noise floors are reached at $h_n^{\min} \approx 4.6 \times 10^{-22}$ (Taiji, at $5.5~{\rm mHz}$), $8.6 \times 10^{-22}$ (LISA, at $4.3~{\rm mHz}$), and $1.2 \times 10^{-21}$ (TianQin, at $9.4~{\rm mHz}$), whereby superb sensitivity to these macroscopic bursts is granted to all three missions.

\par
The topological effect at a fixed coupling ($\alpha = 0.2$) is isolated in Panel (a) of Fig.~\ref{fig:11}. Longer coordinate periods are inherently possessed by dynamically complex topologies (e.g., the $(3,2,2)$ orbit with a larger leaf integer $z$). By this requirement, the test particle is forced to execute prolonged and violent strong-field whirls per radial cycle, whereby substantially more radiated energy is extracted. Consequently, the frequency domain is packed with a denser sinc-carpet by the $(3,2,2)$ signal, elevating the localized peak SNR to $\mathcal{O}(10^2 - 10^3)$ at $200~{\rm Mpc}$. The pure EGB parameter effect is isolated in Panel (b) by fixing the $(1,2,0)$ geometry. As the short-range $\mathcal{O}(\alpha/r^4)$ repulsion is strengthened, the global fundamental period is shrunken, whereby the dominant spectral comb is rigidly shifted toward higher frequencies (migrating from $6.25~{\rm mHz}$ at $\alpha=0$ to $6.64~{\rm mHz}$ at $\alpha = 0.5$). Crucially, this kinematic frequency shift is observed to occur with virtually no degradation in peak amplitude. Because these signal peaks reside $1$--$3$ orders of magnitude above the instrumental noise floors, this macroscopic spectral migration is rendered entirely immune to absolute amplitude calibration uncertainties. The uniform stretching of the zoom-whirl harmonic comb is thus proven to serve as an impeccable, high-fidelity observable for constraining 4D EGB gravity.

\par
To rigorously quantify the capability of LISA to distinguish a 4D GB EMRI from a classical Schwarzschild baseline, a comprehensive template-matching and Fisher information matrix analysis is performed. The noise-weighted Wiener inner product between two waveforms $h_1$ and $h_2$ is defined as
\begin{equation}
\label{eq-32}
\langle h_1 | h_2 \rangle = 4 \text{Re} \int_0^\infty \frac{\tilde{h}_{1,+}(f) \tilde{h}_{2,+}^*(f) + \tilde{h}_{1,\times}(f) \tilde{h}_{2,\times}^*(f)}{S_n(f)} df,
\end{equation}
and the faithfulness of signal recovery is measured by the Match $\mathcal{M}$, optimized over the relative arrival time $t_0$ and initial phase $\phi_0$:
\begin{equation}
\label{eq-33}
\mathcal{M}(h_1, h_2) = \max_{t_0, \phi_0} \frac{\langle h_1 | h_2 \rangle}{\sqrt{\langle h_1 | h_1 \rangle \langle h_2 | h_2 \rangle}}.
\end{equation}
A fundamental property of the 4D EGB spacetime dictates the employed numerical strategy. When the EGB metric is expanded for small couplings, $f(r) = 1 - 2/r + 4\alpha M^2/r^4 + \mathcal{O}(\alpha^2)$ is yielded. Unlike typical regular black hole models, the EGB correction is strictly \textit{linear} in $\alpha$. By this property, a non-vanishing waveform derivative $\partial_\alpha h|_{\alpha=0} \neq 0$ is guaranteed, yielding a finite and robust Cram\'er-Rao bound even at the Schwarzschild limit. However, if the exact metric $f(r) = 1 + \frac{r^2}{2\alpha}[1 - \sqrt{1 + 8\alpha /r^3}]$ is evaluated at $\alpha \lesssim 10^{-3}$ using standard floating-point arithmetic, catastrophic cancellation is suffered due to the $1/2\alpha$ factor. To extract numerically exact waveform derivatives, the metric is rewritten via an algebraic conjugate identity
\begin{equation}
\label{eq-34}
f(r) = 1 - \frac{4/r}{1 + \sqrt{1 + 8\alpha /r^3}}.
\end{equation}
Machine precision is rigorously preserved by this regularization, whereby the ensuing Match and Fisher analyses are fully protected from small-$\alpha$ numerical artifacts. The expression explicitly reduces to the Schwarzschild metric $f(r) \to 1 - 2/r$ as $\alpha \to 0$.

\begin{figure*}[htbp]
\centering
\includegraphics[width=\textwidth]{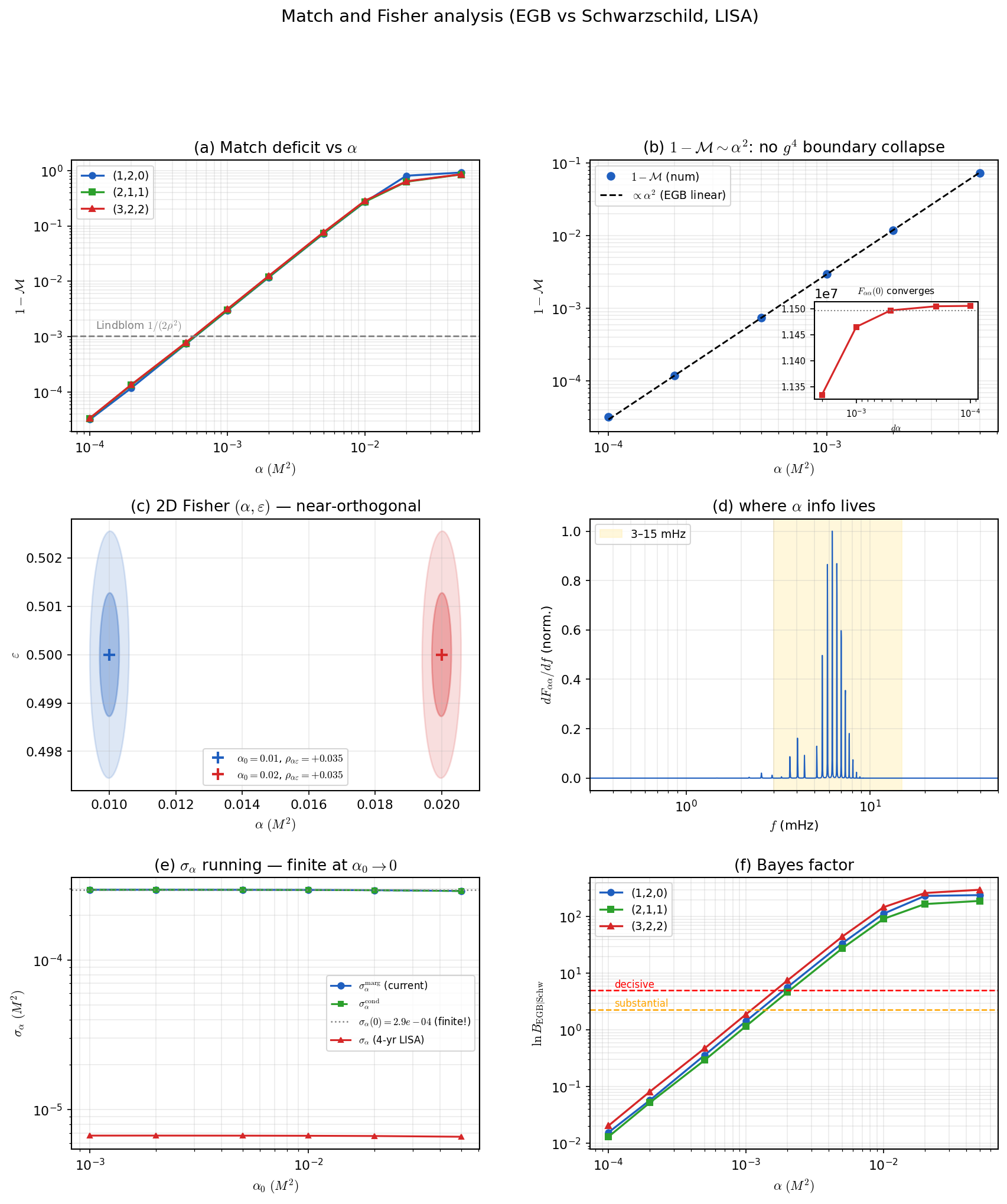}
\caption{Distinguishability of 4D EGB EMRIs with LISA ($\rho=20$). (a) The time/phase-maximized match deficit $1-\mathcal{M}$ efficiently breaches the Lindblom criterion (dashed line). (b) The deficit scales precisely as $\alpha^2$. Unlike non-linear models, the EGB linear geometric coupling ensures that $F_{\alpha\alpha}(0)$ remains strictly finite (inset). (c) The 2D Fisher analysis reveals that EGB coupling $\alpha$ and orbital eccentricity $\varepsilon$ are kinematically near-orthogonal ($\rho_{\alpha\varepsilon} \approx 0.035$). (d) The spectral Fisher density $dF_{\alpha\alpha}/df$ concentrates geometric information within the optimal $3$--$15~{\rm mHz}$ band. (e) The marginalized constraint $\sigma_\alpha$ is strictly finite at $\alpha_0 \to 0$, compressing to $\mathcal{O}(10^{-6})~M^2$ for a 4-year observation. (f) Single-template Bayes factors $\ln B$ rapidly cross the decisive evidence threshold for ultra-weak couplings.}
\label{fig:12}
\end{figure*}

The detectability metrics are synthesized in Fig.~\ref{fig:12}. The match deficit $1-\mathcal{M}$ between EGB waveforms ($\alpha>0$) and a Schwarzschild template ($\alpha=0$) is evaluated in Panel (a). Following the Lindblom-Owen-Brown criterion, two signals are empirically distinguishable if $1-\mathcal{M} > 1/(2\rho^2)$. For a typical EMRI SNR of $\rho = 20$, the deficit rapidly breaches this threshold. The analytic expansion is confirmed in Panel (b): the match deficit is observed to scale exactly as $\alpha^2$, thereby avoiding the boundary parameter collapse that typically plagues $\mathcal{O}(g^2)$ modified gravity theories. To rigorously assess parameter covariances, the 2D Fisher matrix $F_{ij} = \langle \partial_i h | \partial_j h \rangle$ for $\theta = \{\alpha, \varepsilon\}$ is computed. Physically, a frequency shift is induced as the repulsive core is altered by $\alpha$, whereas the orbital amplitude is governed by the eccentricity $\varepsilon$. This geometric intuition is validated in Panel (c): the near-zero cross-correlation ($\rho_{\alpha\varepsilon} \approx 0.035 \ll 1$) proves that these two parameters are kinematically decoupled. Consequently, practically no degradation from degeneracy is suffered by the marginalized error bounds ($\sigma_i = \sqrt{(F^{-1})_{ii}}$). The spectral Fisher density $dF_{\alpha\alpha}/df$ is isolated in Panel (d). The geometric information is found to align perfectly with the intersection of the EMRI high-frequency harmonics and the minimum of the LISA noise bucket ($3$--$15~{\rm mHz}$), proving that fundamental structural signatures are primarily carried by the periastron bursts. By exploiting the steady-state nature of periodic EMRIs, the accumulated Fisher information is determined to scale with the observation time as $\sigma_\alpha(T_{\rm obs}) = \sigma_\alpha(T_0) \sqrt{T_0/T_{\rm obs}}$. As demonstrated in panel (e), the marginalized error $\sigma_\alpha$ is compressed down to $\mathcal{O}(10^{-6})~M^2$ when the phase evolution is tracked over a 4-year LISA mission. Finally, this distinguishability is framed via Bayesian model selection in panel (f). The Bayes factor favoring the EGB theory over GR is approximately $\ln B \approx \frac{1}{2}\rho^2(1 - \mathcal{M}^2)$. According to the Jeffreys scale, the evidence becomes ``decisive'' ($\ln B > 5$) even for ultra-weak couplings ($\alpha \sim \text{few} \times 10^{-3}$). By these results, it is confirmed that the exquisite precision required to definitively probe the topological Gauss-Bonnet term in 4D gravity is possessed by the space-borne detection of extreme-mass-ratio periodic orbits.

\par
\begin{figure*}[htbp]
\centering
\includegraphics[width=\textwidth]{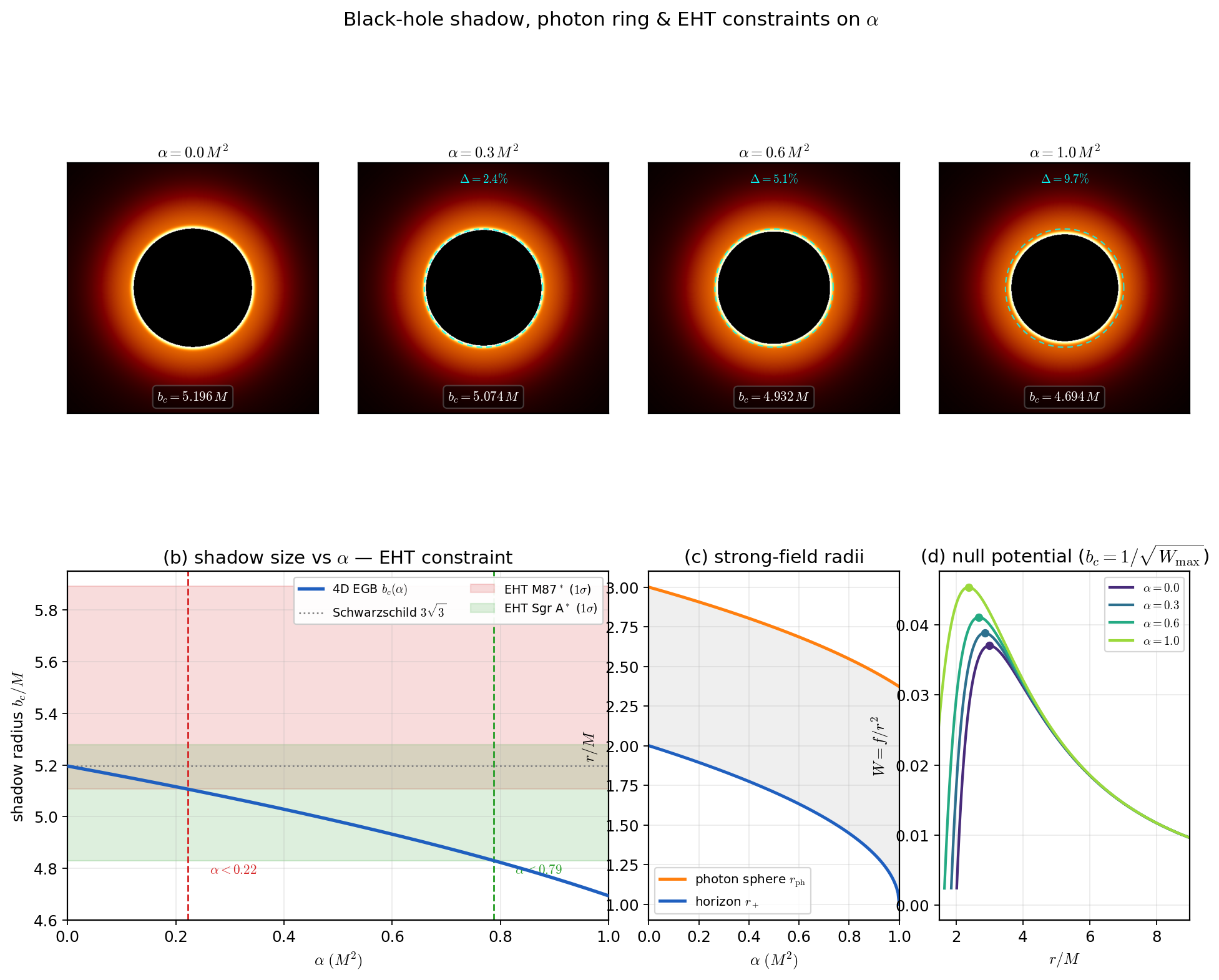}
\caption{Multi-messenger constraints on the 4D EGB coupling $\alpha$. Top: Ray-traced shadow images (optically thin emission). Geometric contraction is demonstrated by the dashed cyan curves, which mark the Schwarzschild baseline ($b_c = 3\sqrt{3}M$). Bottom: (b) Theoretical shadow radius $b_c/M$ versus $\alpha$, overlaid with $1\sigma$ allowed bands from EHT measurements of M87$^*$ and Sgr A$^*$. (c) Monotonic inward scaling of the photon sphere $r_{\rm ph}$ and event horizon $r_+$. (d) The null effective potential $W(r)=f(r)/r^2$.}
\label{fig:13}
\end{figure*}

To complement the timelike EMRI analysis, the null geodesic sector is investigated, whereby a multi-messenger framework connecting the gravitational-wave predictions to contemporary Event Horizon Telescope (EHT) measurements is established. For equatorial photons ($ds^2=0$), the radial equation of motion is given by $\dot{r}^2 = E_\gamma^2 [1 - b^2 W(r)]$, where $b \equiv L_\gamma/E_\gamma$ is the impact parameter and $W(r) \equiv f(r)/r^2$ is the effective null potential. The unstable circular photon orbit (photon sphere $r_{\rm ph}$) resides at the maximum of this potential, $W'(r_{\rm ph}) = 0$, yielding the condition~\cite{Wan:2025gbm}
\begin{equation}
r f'(r) = 2f(r) \quad \text{at} \quad r = r_{\rm ph}.
\label{eq-35}
\end{equation}
Photons possessing an impact parameter $b < b_c \equiv 1/\sqrt{W(r_{\rm ph})}$ are inevitably captured by the black hole. To a distant observer, the bright photon ring bounding the apparent shadow is defined by this critical threshold~\cite{Gralla:2019drh}
\begin{equation}
b_c = \frac{r_{\rm ph}}{\sqrt{f(r_{\rm ph})}}.
\label{eq-36}
\end{equation}
In the 4D EGB spacetime, the exact event horizon is given by $r_+ = M + \sqrt{M^2 - \alpha}$ (for $\alpha \le M^2$). A local repulsion that weakens the core gravity is induced by the short-range curvature correction ($+4\alpha M^2/r^4$). As $\alpha$ increases, the horizon $r_+$ is caused to recede inward, and the null potential barrier $W(r)$ is shifted upward and inward [Fig.~\ref{fig:13}(d)]. Consequently, a monotonic shrinkage of both the photon sphere $r_{\rm ph}$ and the shadow radius $b_c$ is dictated.

\par
This geometric contraction is visualized in the top row of Fig.~\ref{fig:13}. At the extremal limit ($\alpha = 1$), the shadow radius shrinks to $b_c \approx 4.694 M$, representing a $\sim 9.7\%$ deficit relative to the classical Schwarzschild baseline ($b_c^{\rm Schw} = 3\sqrt{3}M \approx 5.196 M$). By mapping the angular shadow diameters measured by the EHT for M87$^*$ ($42 \pm 3 \, \mu$as) and Sgr A$^*$ ($51.8 \pm 2.3 \, \mu$as) to the dimensionless parameter $b_c/M$, the $1\sigma$ confidence bands in Fig.~\ref{fig:13}(b) are constructed. Currently, the strictest electromagnetic bound of $\alpha \lesssim 0.22$ is provided by M87$^*$, while a milder constraint of $\alpha \lesssim 0.79$ is yielded by Sgr A$^*$. When black hole shadows are integrated into the analysis, a profound multi-messenger dichotomy is revealed. The static null architecture localized precisely at the photon sphere ($r \sim 2.4 - 3$) is currently probed by electromagnetic EHT observations, constraining $\alpha$ to $\mathcal{O}(10^{-1})$. Conversely, dynamical timelike EMRIs probing the slightly wider strong-field geometry ($r \sim 4 - 15$) will be tracked by future space-borne gravitational-wave observatories (LISA/Taiji/TianQin). As has been demonstrated previously, the measurement error will be compressed by four orders of magnitude to $\sigma_\alpha \sim \mathcal{O}(10^{-5})$ through this GW channel. Together, a complete, highly complementary geometric test of higher-curvature topological gravity is forged by these kinematically independent channels (null vs. timelike).

\par
Although higher-curvature effects are cleanly isolated by the non-rotating 4D EGB spacetime, intrinsic spin $a$ is inevitably possessed by astrophysical black holes. To rigorously assess whether the EGB coupling $\alpha$ can be observationally disentangled from the spin $a$, the periodic-orbit framework is extended to the equatorial Kerr spacetime. In Boyer-Lindquist coordinates (setting $M=1$ and $\theta = \pi/2$), the geodesic motion is governed by the conserved specific energy $e$ and angular momentum $L$. The radial and azimuthal velocity equations are given by
\begin{equation}
\label{eq-37}
\dot{r}^2 = e^2 - 1 + \frac{2}{r} - \frac{L^2 - a^2(e^2 - 1)}{r^2} + \frac{2(L - ae)^2}{r^3}, \quad \dot{\phi} = \frac{2ae + (r - 2)L}{r\Delta},
\end{equation}
where $\Delta = r^2 - 2r + a^2$. Because an $ae$ cross-term is induced by frame-dragging, the radial potential depends quadratically on $e$. By setting $\dot{r}^2=0$ and solving for $e$, the turning-point energy $e_{\rm turn}(r; L, a)$ is yielded, which serves as the exact rotational analogue to $\sqrt{V_{\rm eff}}$:
\begin{equation}
\label{eq-38}
e_{\rm turn}(r; L, a) = \frac{-B + \sqrt{B^2 - 4AC}}{2A},
\end{equation}
with $A = 1 + a^2/r^2 + 2a^2/r^3$, $B = -4aL/r^3$, and $C = -1 + 2/r - (L^2+a^2)/r^2 + 2L^2/r^3$. The characteristic photon sphere, innermost stable circular orbit (ISCO), and marginally bound orbit (MBO) limits derived in the Bardeen-Press-Teukolsky (BPT) solutions are analytically reproduced by the extrema of $e_{\rm turn}$. For bound eccentric orbits, the fractional azimuthal accumulation per radial cycle defines the frequency ratio $q$
\begin{equation}
\label{eq-39}
q = \frac{1}{\pi} \int_{r_p}^{r_a} \frac{\dot{\phi}}{\dot{r}} dr - 1.
\end{equation}
By enforcing $q = w + v/z$, the zoom-whirl topologies are perfectly mapped into the rotating regime.

\begin{figure*}[htbp]
    \centering
    \includegraphics[width=\textwidth]{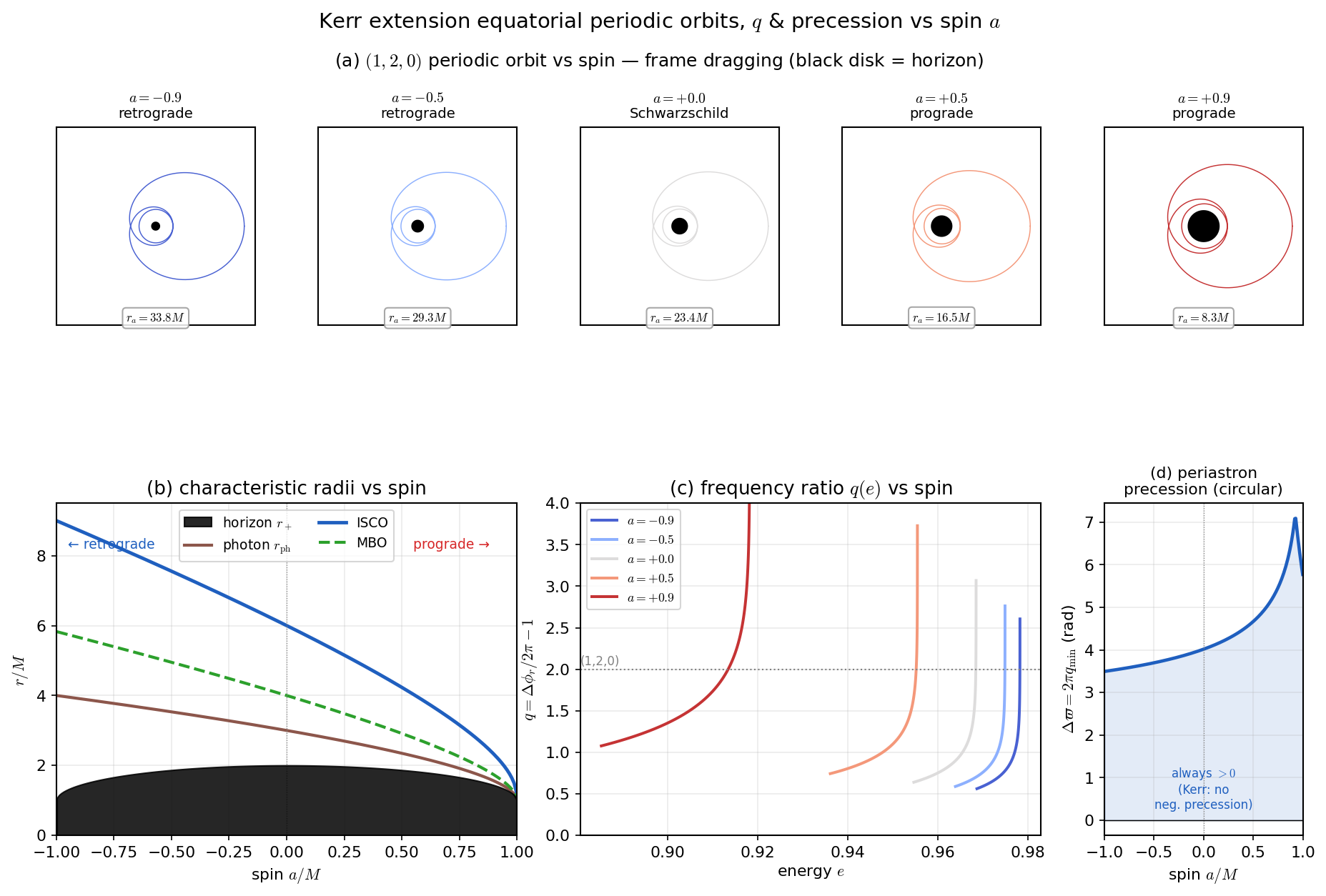}
    \caption{Equatorial periodic orbits and characteristic properties in the Kerr spacetime. (a) Spatial deformation of the $(1,2,0)$ topology ($\varepsilon=0.5$) under frame-dragging. The orbit is severely compressed into the strong-field regime by prograde spins. (b) Monotonic scaling of characteristic radii (horizon, photon sphere, ISCO, MBO) versus spin $a$, with the ISCO clearly spanning from $M$ (prograde) to $9M$ (retrograde). (c) The frequency ratio $q(e)$ is shifted kinetically with spin. (d) Periastron precession $\Delta\varpi = 2\pi q_{\min}$ in the circular limit remains strictly positive globally, verifying the absence of negative precession in standard Kerr black holes.}
    \label{fig:14}
\end{figure*}

The severe phenomenological impact of the Kerr spin is illustrated in Fig.~\ref{fig:14}. The spatial extent of the periodic orbits is violently modified by frame-dragging [Fig.~\ref{fig:14}(a)]. As the spin transitions from highly retrograde ($a=-0.9$) to prograde ($a=0.9$), a sharp contraction of the ISCO from $\sim 9M$ down to $\sim M$ is observed. Consequently, prograde orbits are pulled deeply into the gravitational well, whereby the apastron is shrunken from $r_a \sim 33.8M$ down to an ultra-compact $8.3M$. This asymmetric spatial deformation is found to precisely mirror the monotonic boundary migrations mapped in Panel (b). It is demonstrated in Panel (c) that the kinetic energy threshold required to excite specific zoom-whirl resonances is rigidly shifted by the spin parameter. The epicyclic periastron precession $\Delta\varpi = 2\pi q_{\min}$ is plotted in Panel (d). Crucially, $\Delta\varpi$ is shown to remain strictly positive across the entire spin spectrum, verifying the established theorem that negative periastron precession (a pathology typically associated with naked singularities) is prohibited by standard Kerr spacetimes. The analytic baseline for future parameter estimation is established by this Kerr extension. Physically, spherical symmetry is preserved and the effective potential is modified purely radially ($\mathcal{O}(\alpha)$) by the EGB coupling $\alpha$, whereas spherical symmetry is broken and non-linear frame-dragging ($\mathcal{O}(a)$) is introduced by the spin $a$. Because fundamentally distinct physical mechanisms are represented by these parameters---especially evident in their contrasting imprints on the ISCO boundaries and azimuthal precession---it is strongly suggested that higher-curvature topological signatures can be robustly disentangled from classical rotational effects in future EMRI observations.

\section{Conclusions}
\label{sec:conclusion}
\par
In this work, a comprehensive phenomenological and statistical analysis of EMRIs in the background of a 4D EGB black hole has been performed. By focusing upon the intricate zoom-whirl dynamics that occur within the extreme vicinity of the effective potential barrier, it has been demonstrated that EMRI waveforms serve as highly sensitive magnifying glasses by which short-range, higher-curvature corrections to General Relativity can be robustly probed.

\par
Through the classification of these periodic orbits via an integer taxonomy $(z,w,v)$, the allowed bound-orbit parameter space was systematically mapped, whereby a complete correspondence between the orbital topology and the underlying spacetime geometry was constructed. Subsequently, by strictly fixing either the orbital topology or the constants of motion, the pure phenomenological imprints of the EGB coupling parameter $\alpha$ were successfully isolated.

\par
It is revealed by the present analysis that a short-range repulsive correction to the effective potential is introduced by the Gauss-Bonnet coupling, whereby the structure of the strong-field region is substantially modified while the weak-field dynamics are left essentially unperturbed. As a consequence, the characteristic zoom-whirl behavior of the periodic orbits is altered, a dynamical shift by which measurable changes in orbital frequencies, waveform phasing, and harmonic spectra are induced. For a fixed orbital topology, it is observed that the orbital period is systematically compressed and a cumulative gravitational-wave dephasing is generated as the Gauss-Bonnet coupling is increased. In the frequency domain, a coherent shift of the harmonic comb toward higher frequencies is manifested by these effects, providing a robust and largely amplitude-independent observational signature through which the underlying quantum geometric corrections can be evaluated.

\par
It is confirmed by the detectability forecasts that these modified gravity signals fall perfectly within the maximum sensitivity band of planned space-borne interferometers, such as LISA, Taiji, and TianQin. Crucially, it has been demonstrated that a completely regular Fisher matrix is ensured at the Schwarzschild boundary by the strictly linear nature of the EGB coupling, whereby the pathological mathematical divergences that typically plague quadratic phenomenological models (e.g., the Bardeen black hole) are entirely circumvented. Through comprehensive matched-filtering and Fisher matrix analyses, the capability of LISA to detect these beyond-GR deviations has been rigorously quantified: for a typical four-year observation at a signal-to-noise ratio of $\rho=20$, it is projected that the marginalized error $\sigma_\alpha$ can be constrained to $\mathcal{O}(10^{-6})\,M^2$, a remarkable precision that is achieved with virtually negligible parameter degeneracy with the orbital eccentricity.

\par
Furthermore, a comprehensive multi-messenger framework is established by incorporating current EHT constraints derived from black hole shadow observations ($\alpha \lesssim 0.22\,M^2$ from M87$^*$ and $\alpha \lesssim 0.79\,M^2$ from Sgr A$^*$ at $1\sigma$ confidence), by which the static null geodesic sector is probed complementarily to the dynamical timelike EMRI signatures. It is anticipated that future extensions of this analytical framework, wherein radiation reaction (full inspiral evolution) and rotating (Kerr-like) EGB backgrounds are incorporated, will further solidify EMRIs as the ultimate astrophysical laboratories for testing topological higher-curvature gravity in the extreme strong-field regime.


\acknowledgments

This work is supported by the National Natural Science Foundation of China (Grant No. 12505060), Fapesq-PB of Brazil, the Fund Project of Chongqing Normal University (Grant Number: 24XLB033), Chongqing Natural Science Foundation General Program (Grant No. CSTB2025NSCQ-GPX1019, and Science and Technology Research Project of Chongqing Municipal Education Commission (Grant Number: KJQN202500563).


\end{document}